\documentclass[aip,cha,reprint,showpacs,showkeys,twocolumn,floatfix]{revtex4-1}
\usepackage{amssymb}
\usepackage{amsmath}
\usepackage{amsfonts}
\usepackage{graphicx}
\usepackage{epstopdf}%\usepackage{color}

\newcommand{\Z}{{\mathbb{Z}}}

\begin{document}
%\nocite{eam}
%\bibliographystyle{elsarticle-harv}
%\bibliographystyle{apsrev}  %for prl style
%\bibliographystyle{unsrt} %for long titles
%\bibliographystyle{apsrmp}
%\bibliographystyle{abbrv}% no abrevia, to be used
%\bibliographystyle{alpha}
%\bibliographystyle{elsarticle-num}

%\newcommand{\C}{$^\circ$C}
%\newcommand{\Cmath}{^\circ\text{C}}
%\newcommand{\juan}[1]{{\bf Juan: #1}}  %writes in bf with Juan in front of the text
%\newcommand{\juan}[1]{{#1}} %uncomment this command and comment the previous one to eliminate the Juan and bf.

\title{Nonlinear waves in a model for silicate layers}%

\author{Juan F. R. Archilla}
\email[]{archilla@us.es}
\affiliation{Grupo de F\'{\i}sica No Lineal,  Universidad de
Sevilla, ETSI Inform\'{a}tica,
  Avda Reina Mercedes s/n, 41012-Sevilla, Spain}
\author{Yaroslav Zolotaryuk}
\affiliation{Bogolyubov Institute for Theoretical Physics,
  National Academy of Sciences of Ukraine,   vul. Metrologichna 14-B
  03680 Kiev, Ukraine}
\author{Yuriy A. Kosevich}
\affiliation{Semenov Institute of Chemical Physics, Russian
Academy of Sciences,  Kosygin street 4, 119991 Moscow, Russia}
\author{Yusuke Doi}
\affiliation{Department of Adaptive Machine Systems, Graduate School of Engineering,Osaka University, 2-1 Yamadaoka, Suita, Osaka 565-0871, Japan}
\date{\today}

 \begin{abstract} Some layered silicates are composed of positive ions, surrounded by layers of ions with opposite sign.  Mica muscovite is a particularly interesting material, because there exist fossil and experimental evidence for  nonlinear excitations transporting localized energy and charge along the cation rows within the potassium layers.  This evidence suggest that there are different kinds of excitations with different energies and properties. Some of the authors proposed recently a one-dimensional model based in physical principles and the silicate structure. The main characteristic of the model is that it has a hard substrate potential and two different repulsion terms, between ions and nuclei.  In a previous work with this model,  it was found the propagation of crowdions, i.e., lattice kinks in a lattice with substrate potential that transport mass and charge. They have a single specific velocity  and energy coherent with the experimental data. In the present work we perform a much more thorough search for nonlinear excitations in the same model using the pseudospectral method to obtain exact nanopteron solutions, which are single kinks with tails, crowdions and bi-crowdions. We analyze their velocities, energies and stability or instability and the possible reasons for the latter. We relate the different excitations with their possible origin from recoils from different beta decays and with the fossil tracks.  We  explore the consequences of some variation of the physical parameters because their values are not perfectly known. Through a different method, we also have found stationary and moving breathers, that is, localized nonlinear excitations with an internal vibration. Moving breathers have small amplitude and energy, which is also coherent with the fossil evidence.
 \end{abstract}
\maketitle

\newcommand{\juan}[1]{{\bf\em  JUAN: #1}}
\newcommand{\yuriy}[1]{{\bf YURIY: #1}}
\newcommand{\yaroslav}[1]{{\bf YAROSLAV: #1}}
%\newcommand{\juan}[1]{{}} \newcommand{\yuriy}[1]{} \newcommand{\yaroslav}[1]{}

%\markboth{JFR Archilla, Y Zolotaryuk, YuA. Kosevich}
%{Kinks, double kinks, nanopterons and breathers in a model for layered silicates}

%\maketitle

{\bf
Some fossil dark tracks in a layered silicate, mica muscovite, have been identified with localized excitations that travel through the rows of atoms within layers. Also, experiments show that these excitations are able to expel atoms from the surface of the mineral  and transport charge. The different tracks that can be observed suggest that different kinds of excitations exist. In this paper we use a model where the potentials have been obtained from the physical properties and the structure of the mineral. We make a systematic search for  the possible nonlinear excitations and find different ones. Some imply the movement of charge and mass, others have smaller energy and are only vibrational. Some are localized, but others are coupled to low amplitude plane waves. We explore the energies, velocities and stability of these solutions and explain how they can  be produced by the recoil after different radioactive decays of the isotope $^{40}$K and relate them with the different tracks.
}
\section{Introduction}

Muscovite is a layered silicate that has been from long ago of interest for different reasons. It was the second material\,\cite{silkbarnes59} for which nuclear tracks from fission fragments were observed, the first material for which fossil nuclear tracks were discovered\,\cite{pricewalker62}. Also fossil tracks from different elementary swift particles as positrons and muons could be observed\,\cite{russell-prb1967}. These tracks were formed by the precipitation of excess impurity of Fe growing by accretion into the mineral magnetite and can be viewed without instruments. Many other tracks were in the closest-packed directions of the K$^+$ plane and were attributed to some kind of quasi one-dimensional lattice excitations, whose exact nature is not well known. See Ref.\,\cite{russell-tracks2015book} for a recent review. The existence of these excitations was also demonstrated experimentally by bombarding with $\alpha$ particles a muscovite sample and detecting the ejection of an atom at the direction of the lattice closest-packed lines\,\cite{russell-experiment2007}. Recently, it was deduced that the lattice-excitations, called quodons, could be of many different types with different charge states\,\cite{archillaLoM2016}, and demonstrated experimentally\,\cite{russell-archilla2017}.  Trying to understand the basic properties of the mineral that led to the existence of nonlinear excitations, a one-dimensional model was introduced with potentials deduced from basic principles and physical properties that led to the numerical finding of the existence of a moving kink or crowdion, that is, a moving interstitial, with a single energy which was the appropriate for the experimental and fossil observations\,\cite{archilla-kosevich-pre2015,archilla-kosevich2015b-book}.

In this article we use powerful numerical techniques to perform a more systematic search of nonlinear waves in the same model. We have found other solitary waves as kinks, double kinks and nanopterons with different energies and velocities. However, among them only the faster entities as one single kink, two double kinks and fast nanopterons are stable. Their approximate velocity  is about 5 times the sound velocity in the system. We have also found stationary breathers that cannot be moved, and also small amplitude moving breathers that propagate smoothly through the lattice. The physical consequences of these findings are analyzed.

The generic solutions found are nanopterons, i..e, non-local solutions that can be described as the coupling of a kink and a plane wave. Nanopterons can be considered as an approximation to a physical solution, but can also be of interest by themselves\,\cite{boyd1990-non,shiroky2018,vorotnikov2018,hoffman2017,kim2015}. Nonlocal breather solutions are also subject of current research\,\cite{yoshimura2007,sato2015,doi2016}.

It should be stressed that finding localized non-radiating
travelling-wave solutions in lattices with an on-site potential was and remains an important problem
in the physics of nonlinear waves. Discreteness of the media
seriously obstructs existence of exact moving topological solitons
\cite{pk84pd}. However, it has been shown in numerous cases both
analytically \cite{s79prb,fzk99pre,opb06n,dkksh08pre} and numerically
\cite{zolotaryuk97,sze00pd,kzcz02pre,acr03pd} that these solutions exist, also in two-dimensional lattices \cite{bajars-springer2015,bajars-physicad2015,dmitriev18},%bajars-springer-article2015
despite the obvious resonance with the plane waves of the system.
Therefore they fall into the family of the {\it embedded} solitons, that
are literally embedded in the linear spectrum of the
system as defined in Ref.\,\cite{zolotaryuk-cmyk01pd}. Contrary to the
conventional solitons that usually exist as a one
family of solutions, where the parameter may be their velocity, frequency
or some other characteristic, the embedded solitons form a discrete
set of solutions. In this paper we
are going to demonstrate the existence of nonradiating moving topological
solitons in a model for mica muscovite with realistic potentials.

The outline of the article is as follows: we describe the mineral and previous results in Sect.~\ref{sec:muscovite} and the details of the model in Sect.~\ref{sec:model}. Section~\ref{sec:phonons} describes the peculiarity of the phonon spectrum and the phonon different velocities. Section.~\ref{sec:method} explains the pseudo-spectral method for calculating travelling solitary waves with precision. In Sect.~\ref{sec:results}, we describe the characteristics of the nanopterons and kinks obtained, their velocities and energies. Their  stability is analyzed in Sect.~\ref{sec:stabilityQ_1}.  In Sect.~\ref{sec:otherparameters} we explore the consequences of changing the strength of the interaction or the on-site potential. In Sect.~\ref{sec:double} the properties of some double kinks or 2-crowdions are presented. Breathers are studied in Sect.~\ref{sec:breathers} and the article finishes with the conclusions in Sect.~\ref{sec:conclusions}.

\section{Muscovite and nonlinear excitations}
\label{sec:muscovite}
In this section we describe the silicate structure of muscovite, the physics of the model, previous results, and some considerations about the possible existence of different nonlinear excitations in the material and the model.

\subsection{Silicate structure}
Most silicates are formed by the basic building block of a tetrahedron formed by the cation silicon surrounded by oxygen anions at the vertices. Often those oxygen anions are shared with other tetrahedra forming sheets of several layers, sometimes tetrahedra sheets, joined with other ones or with octahedra sheets, are sandwiched between tetrahedra sheets bringing about thicker layers. These sheets can be joined together by  van der Waals forces. Often, substitutions of silicon by other cations like aluminium with different charge are common and therefore the sheets have a net electrical charge. Ions of different sign are therefore brought in to neutralize the charge and the forces between sheets become stronger and predominantly electrostatic.  The result is a layer of ions that are coupled to the silicate sheets but less stronger that the silicate structure of the sheets. The inter-layer ions are kept in place at specific places where the attraction of the silicate layer is stronger. If we consider the ion lattice, it can be viewed as subjected to an on-site potential where the ions interact between them through electrical repulsion. It is a natural and important example where on-site potentials become important in determining the properties of a lattice.
The layered silicate mica muscovite is of particular interest in this article. It consists of a silicate layer with the structure of tetrahedra-octahedra-tetrahedra. Substitution of Si$^{+4}$ by Al$^{+3}$ results in a net negative electrical charge, which is compensated by the introduction of a cation layer made out of K$^+$. Often, there are also  different substitutional impurities which have different properties, however we will refer in this article to the ideal structure.

\subsection{A realistic model for muscovite and the existence of a crowdion}
A model with realistic potentials was constructed starting from the simplest model with only electrostatic repulsion between K$^+$ ions\,\cite{archilla_ujp} and later augmented with a Ziegler-Biersack-Littmark (ZBL) short-range repulsive potential between K$^+$ ions\,\cite{ziegler2008}, which  was important for describing the interaction at very short distances bringing about reasonable minimal distances for energetic excitations\,\cite{archilla-kosevich-pre2015,archilla-kosevich2015b-book}. In the latter references, it was also introduced an on-site potential calculated using empirical potentials\,\cite{gedeon2002} and the interaction with the two closest two-dimensional layers of ions above and below the K$^+$ layers with some adjustment to take into account further layers. The resulting periodic potential, which we will use in this article, could be expressed as a cosine Fourier series with five terms and was able to reproduce approximately the absorption infrared frequency 110\,cm$^{-1}$ obtained experimentally\,\cite{diaz2000} and the 20\,eV potential well obtained with molecular dynamics\,\cite{CollinsAM92}. This model led to the numerical finding of a crowdion, that is a kink in a lattice with a periodic on-site potential, where a particle is moved to another site pushing the particle that sits there and becoming a replacement cascade, which can also be described as a moving interstitial~\cite{archilla-kosevich-pre2015,archilla-kosevich2015b-book}. That crowdion or lattice kink has several important characteristics from the mathematical and physical point of view: a) it has a unique supersonic velocity and energy and travels long distances; b) it survives to perturbations of the media as nonlinear waves and thermalization\,\cite{archilla-kosevich2015b-book}; c) its energy of about 26 eV is in the right order of magnitude for mica muscovite crystal, smaller than most of the recoils after $\beta$ decay of $^{40}$K\,\cite{radionuclides2012,cameron2004}, which has a maximum recoil of ~42\,eV for $\beta^-$\,\cite{archilla-kosevich2015b-book}, and larger than the typical ejection energies of 4-8\,eV\ of atoms in silicates \cite{kudriavtsev2005}, making a good candidate for the experimental results\,\cite{russell-experiment2007}. The property that supersonic crowdions have a discrete set of velocities is well known \cite{kosevich73,savin95}. The key point for the existence of supersonic crowdions is the existence of the on-site potential, which can be provided by an external system or simply by the rest of the crystal, if we consider for example a row of nearest-neighbour atoms as a subsystem subjected to the lattice potential. In a layered silicate crystal the on-site potential is provided by the silicate sheets above and below the layer of potassium atoms. The high bond stiffness of the silicate layers make them  dynamically decoupled from the K$^+$ layer.
%juan: include some data:
%internal vibrations of Si04 ~ 80 meV~7.7 kjm ~1.9·1013 Hz~644cm-1~927 K
%Heavy atom vibrations
% with translations/rotations of Si04  <30 meV~2.89 kjm~7.24·1012 Hz~241cm-1~350 K
%bond-angle vibrations  0-Si-0:
%~ 30-60 meV~2.9-5.8 kjm~7.24-14.48·1012 Hz~241-482cm-1 ~350-700 K
%band gap
%Stretching vibrations ~ 100 meV ~ 9.6 kjm~2.4·1013 Hz~ 805 cm-1~1160 K
%K+  in the a direction 110cm-1

\subsection{Other nonlinear localized waves in muscovite}
In this paper we expand the search to other nonlinear localized waves in the same model for mica muscovite. The reason for that is partly theoretical, as a peculiar characteristic of the on-site potential is that it is hard for relatively small amplitudes and later becomes soft. This hard-soft characteristic does not appear as a mathematical curiosity but as a consequence of constructing a potential based on physical laws of a particular system. The same can be said about the interaction potential between ions, which is obtained simply by electrostatic repulsion and short range universal ZBL repulsion, which describes the electrostatic repulsion between the nuclei screened by the electron shell. On the other hand the dark tracks in muscovite that are attributed to nonlinear lattice excitations are not equal, for example, some are strong and continuous and other are faint and discontinuous and are scattered at 60$^\circ$ from the strong ones\,\cite{russell2018}. It can  also be deduced from the analysis of the first collisions after $^{40}$K recoil\,\cite{archillaLoM2016}. These secondary tracks cannot have the same energy as the primary one, which means there may be different excitations at play. We describe three types of lattice excitations for this model, kinks (without tails), nanopterons, which are kinks with tails, and breathers, which are vibrational localized oscillations.

The terminology is not uniform partly due to different characteristics of a excitation depending of the variable used. The word soliton is used also for kinks, because the bond compressions and the velocities are solitons, i.e., their amplitudes tend to zero at $\pm\infty$, and breathers can be seen as envelope solitons with an internal vibration. Kinks and crowdions are the same entity in a system with an external potential as the one studied here.

A method for finding kinks was developed in\,\cite{zolotaryuk97} where the terminology bound solitons was used because the profile of the bond compressions looks like the union of several solitons.  Nonlocal kinks or nanopterons that are composed of a localized kink core coupled with a plane wave, which are solutions of the system, are constructed. Among them we search for solutions that have zero amplitude for the plane wave, which are proper kinks.  Nanopterons are not properly kinks because they have oscillating plane wave asymptotics.
They are however long-lived and
appear in many different physical systems\,\cite{boyd1990-non}
and are therefore interesting by themselves.

Breathers are found by continuation from the continuous limit. There are site-centered ones that are  very stable and have a wide range of frequencies. They are however not movable but they can have relevance in other phenomena as the experimentally observed fast reaction kinetics in muscovite\,\cite{archillaJPCB2006,archillaAIP2008,dubinko-archilla2011}.  However, small amplitude moving breathers are found with simple initial conditions.

\subsection{Nonlinear waves in two-dimensional lattices}
\label{ssec:2Dwaves}
There is also interest in the study of the possibility of excitation and propagation of solitons, discrete breathers and kinks in two-dimensional lattices. Different interatomic potentials, including discrete sine-Gordon and Lennard-Jones~\cite{marin2D1998,marinCuprate2001,bajars-physicad2015,bajars-springer2015} and Morse~\cite{chetverikov17,dmitriev18} potentials, were used in these studies which confirm the existence and long lifetime of the kink solutions in two-dimensional lattices, including a two-dimensional model proposed for mica muscovite~\cite{bajars-springer2015,bajars-physicad2015}. In the present work, we consider a one-dimensional lattice model which allows us to perform more rigorous study of the existence and stability of  nonlinear excitations in a system with realistic interatomic potentials. This also allows us to obtain the values of magnitudes as frequency, velocity and energy in physical units.

\section{A model with a hard substrate potential and two repulsive interactions}
\label{sec:model}
As stated above the model and parameters correspond to the system described in detail in Refs.\,\cite{archilla-kosevich-pre2015,archilla-kosevich2015b-book}. We consider a one-dimensional system of particles corresponding to ions K$^+$ separated by a distance $r$ which in equilibrium
is the lattice constant $a=5.19$~\mbox{}\AA\,  which is set to 1 in scaled form with distance unit $u_L=a$. The interaction between particles is given in scaled form as:
\begin{equation}
U(r)=\frac{1}{r}+\frac{B\textrm{e}^{\displaystyle -r/\rho}}{r}\,,
\label{eq:Ur}
\end{equation}
where the first term corresponds to the Coulomb repulsion between ions and the second Yukawa-type potential is the ZBL potential which describes the Coulomb repulsion between nuclei screened by the electron shell\,\cite{ziegler2008}. The actual ZBL potential has four Yukawa terms that are important for different energy ranges from eV to MeV, but  for the range of energies below 200\,KeV one term is enough. The scaled unit of  energy is the Coulomb energy at the equilibrium distance $u_E=K_ce^2/a\simeq 2.77$\,eV, and the constant $B=184.1$ is of the order of $Z^2$, being $Z=19$ the atomic number of K. The screening range is $\rho=0.0569$, which corresponds to $\simeq 0.3$\,\AA\, in physical units.
The on-site potential has the periodicity of the lattice and is given in scaled units by
%--------------------------------------------------------------------
\begin{equation}\label{eq:Vu}
V(u)=\sum_{m=0}^4 V_m \cos(2\pi m u)\,,
\end{equation}
with $\{V_m\}=\gamma [2.4473,-3.3490,1.0997,-0.2302,0.0321]$.
It has been obtained using empirical potentials \cite{gedeon2002} for the interaction with the two-dimensional silicate layers above and below the K$^+$ layers\,\cite{archilla-kosevich-pre2015}. The parameter $\gamma=1$ will be used later to check the effect of a change on the strength of the on-site potential.

Each particle with index $n$ is described by a coordinate $x_n$, which in equilibrium has the value $x_n^0=n$. We perform a change of variables to  $u_n=x_n-n$ which represents the deviation from the equilibrium position. In the variables $u_n$ the interaction potential is given by:
\begin{eqnarray}\label{eq:Un}
\lefteqn{U(u_{n+1}-u_n)=}\nonumber\\&& \frac{1}{1+u_{n+1}-u_n}+%\nonumber\\
\frac{B}{1+u_{n+1}-u_n}e^{-(u_{n+1}-u_n+1)/\rho}\,
\end{eqnarray}
without change for $V(u_n)$ due to its periodicity.

The kinetic energy of each particle is given by
%--------------------------------------------------------------------
\begin{equation}\label{eq:kinetic}
K_n=\frac{1}{2}m_K{\dot u}_n^2\,,
\end{equation}
%--------------------------------------------------------------------
with $m_K$=39.1\,amu being the mass of a potassium atom. It will be taken as the unit of mass $u_M=m_K$ in scaled variables, in which the kinetic energy is simply $\frac{1}{2}\dot{u_n}^2$. The scaled unit of time $u_\tau$ can now be obtained from $u_E=u_M (u_L/u_\tau)^2$, being $u_\tau=(m_k a^3/K_c e^2)^{1/2}\simeq 0.2$\,ps.
Therefore, the total energy is given in scaled units by:
\begin{equation}\label{eq:Htotal}
H=\sum_n {\cal E}_n=\sum_n\left
(\frac{1}{2}{\dot u_n}^2+V(u_n)+U(u_{n}-u_{n-1})\right )\, .
\end{equation}
The equations of motion have the following form:
%--------------------------------------------------------------------
\begin{equation}
{\ddot u}_n=-V'(u_n)+U'(u_{n+1}-u_n)-U'(u_{n}-u_{n-1}),\,
\label{eq:ddotun}
\end{equation}
%--------------------------------------------------------------------
with $n=1,2,\ldots,N$. The periodic boundary conditions
$u_{n+N}=u_n+Q$ are applied with $Q$ being the topological charge that will be explained below.

%#############figure1 #######################################################
%####################################################################
\section{Phonons and their phase and group velocities}
\label{sec:phonons}
\begin{figure}[htb]
\begin{center}
\includegraphics[width=0.8\columnwidth]{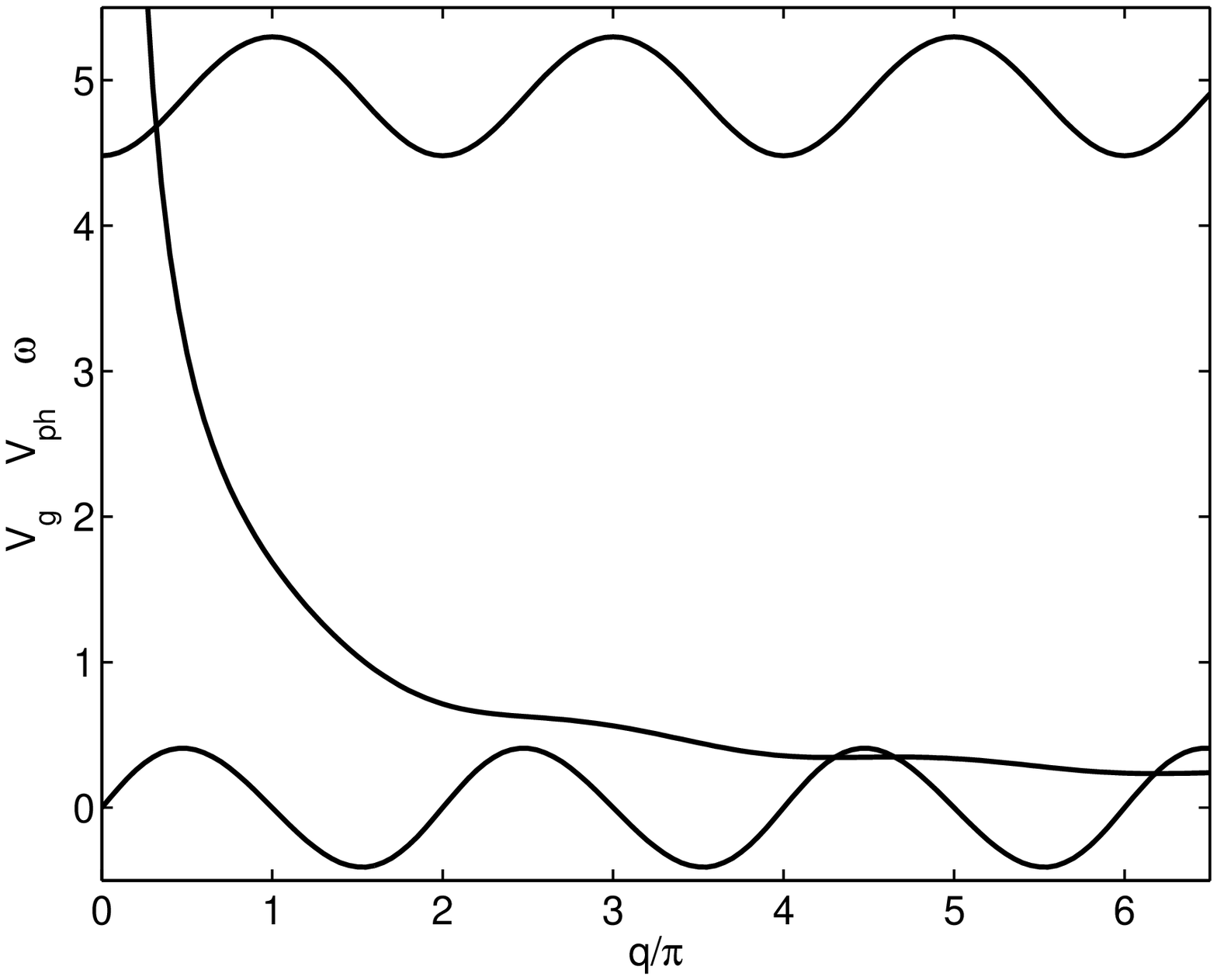}\\
\includegraphics[width=0.8\columnwidth]{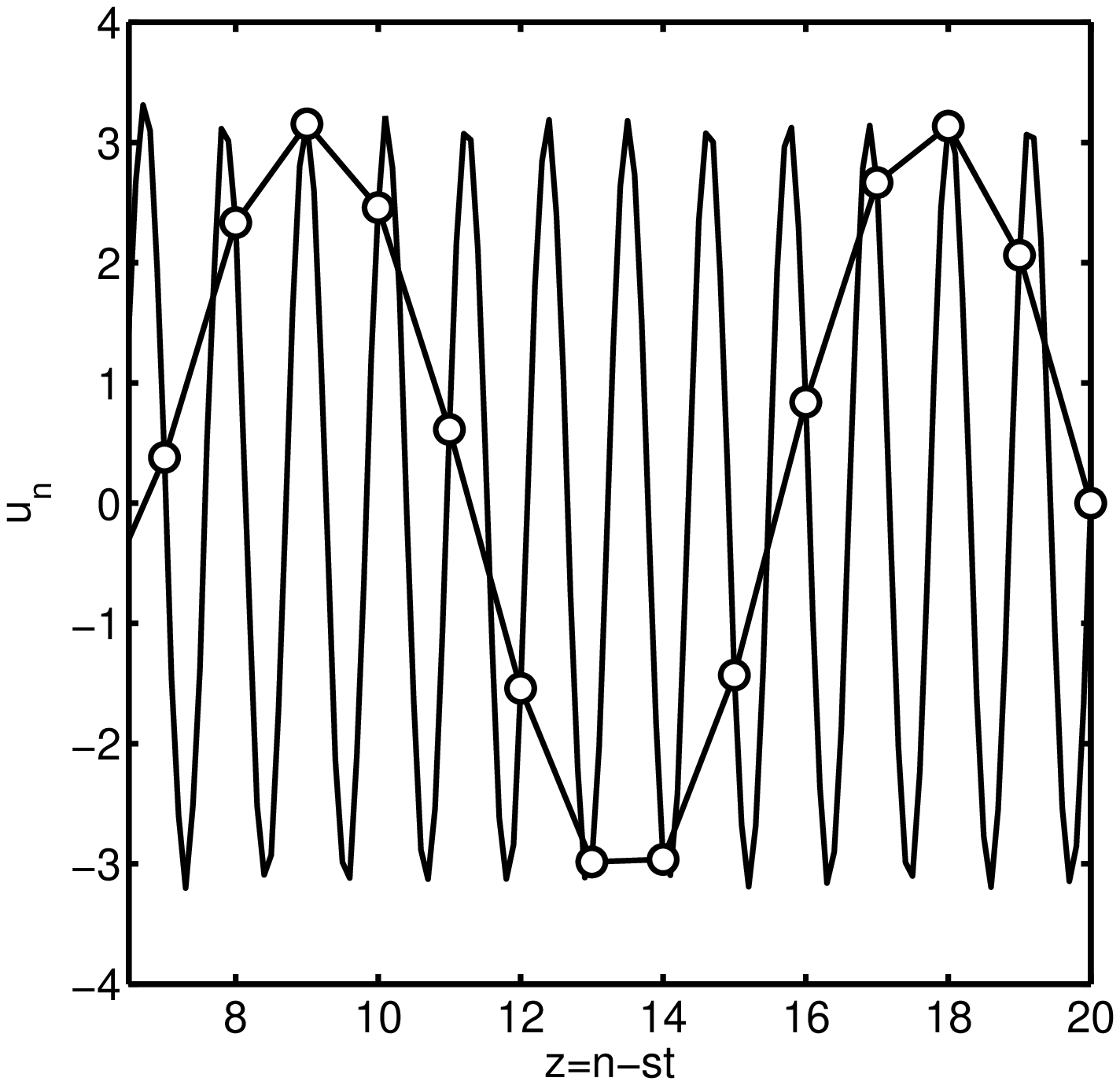}
\end{center}
\caption{{\bf Top:} Phonon properties: dispersion relation $\omega(q)$ (upper oscillating curve), phase velocity $V_\textrm{ph}$ (monotonic function) and group velocity $V_\textrm{g}$ (lower oscillating curve). It can be seen that for any value of the velocity of a solitary wave there is a plane wave with the same phase velocity for which the wavenumber $q$ and
the frequency can be immediately deduced. The wavenumber $q$ can be considered in the extended zone but it also has a representative in the first Brillouin Zone. The group
velocity $V_\textrm{g}$ is zero at $q=0$ and $q=\pi$, having the maximum value $V_\textrm{g}\simeq 0.4091$ at $q=0.4733\pi$. {\bf Bottom:} Plane wave associated with the sliding kink with velocity  $s=0.825$  found in Sect.~\ref{sec:results} where it can be seen the plane wave with phase velocity $V=s$ and $q=1.77\pi$, in the extended zone (smaller wavelength), and its representative in the 1st Brillouin zone, with $q'=q-2\pi=-0.23\pi$ (larger wavelength) and velocity $V'=-6.37$. The minimum velocity in the first Brillouin zone is 1.68.}
\label{fig:phonons}
\end{figure}

The system described by Eq.~(\ref{eq:ddotun}) can be linearized for small amplitudes of $u_n$ and the phonons
properties can be obtained easily\,\cite{archilla-kosevich2015b-book}. They are typical of systems with an on-site potential but some of their properties and the specific values of some magnitudes worth being remarked.
The phonon dispersion is given by:
\begin{eqnarray}
\omega(q)&=&\left [\omega_0^2+4c_s^2\sin^2\left (\frac{q}{2}\right
)\right ]^{1/2}\,,\label{eq:phonondispersion}\\
\mathrm{with}\quad\omega_0^2&=&V''(0)=\gamma\sum_{n=1}^4 (2\pi n)^2V_n\, ,\nonumber
\end{eqnarray}
where $q$ is the phonon wavevector. The frequency $\omega_0=4.48$  corresponds to linear oscillations of the isolated particle and $c_s=\sqrt{2}$
is the sound velocity for the system without substrate, i.e., it is both the group and phase velocity for $q\rightarrow 0$, and it is also the maximum phonon velocity in the system without substrate.
The wavevectors $q$ correspond to plane waves $\cos[q n-\omega(q)t]$ and $\sin[q n-\omega(q)t]$ and due to the periodic boundary conditions they become
discrete variables $q_n=2\pi m/N$, for any integer $m$. As the wavenumbers $q$ and $q+2\pi n$ are equivalent, it is usually reduced to the first
Brillouin zone $q\in[-\pi, \pi]$ or the positive part of it for symmetry. However, the wavevectors in the extended zone are important in this system.
The phonon spectrum is bounded from below by $\omega_0$ and therefore the phase velocity $V_\text{ph}=\omega(q)/q$ tends to $\infty$ when $q\rightarrow 0$.
This is an important property for Klein-Gordon system (systems with substrate potential) very different from FPU systems (without substrate). Note also,
that $V_\text{ph}$ is different for the wavevector $q$ and $q+2\pi$ in spite of representing indistinguishable phonons. This brings about that the phase velocities although
bounded from below in the first Brillouin zone by $V_\text{ph}(\pi)\simeq 1.68$, they are not in the extended zone. They have however a representative in the first Brillouin zone as can be seen in Fig.~\ref{fig:phonons}-top.

However, the group velocity
\begin{equation}
V_\textrm{g}=\frac{\partial\omega}{\partial q}=\frac{c_s^2\sin(q)}{\omega}=\frac{c_s^2\sin(q)}{[\omega_0^2+4c_s^2\sin^2(q/2)]^{1/2}}\,
\label{eq:groupV}
\end{equation}
is zero at $q=0$ and $q=\pm \pi$ and has a maximum value which is easy to calculate $V_\text{g,max}\simeq 0.4091$ for $q=0.4733\pi$.  The phonon dispersion law and the phase and group
velocities can be seen in Fig.~\ref{fig:phonons}.

Therefore, for any velocity of the kinks there is always a plane wave with the same velocity, which can be coupled with it to form a nanopteron, in some cases
the amplitude of the plane wave becomes zero, which corresponds truly to kinks. This is
a different situation with FPU systems where the phonon velocities are bounded from above and supersonic kinks have velocities above the phonon ones.
However, plane waves cannot transport energy as they are extended in the whole space, for that wave groups are necessary. In systems with substrate, 'supersonic' means above the group velocities. In this work we will find that all kinks and nanopterons are supersonic.

%####################################################################
%####################################################################
\section{Pseudo-spectral method for calculating solitary waves}
\label{sec:method}
%####################################################################
%####################################################################
%####################################################################
%####################################################################
%@@@@@@@@@@@@@@@@@@@@@@@@@@@@@@@@@@@@@@@@@@@@@@@@@@@@@@@@@@@@@@@@@@@@@@
%
%-------------------Figure2 ------------------------------------------
\begin{figure*}[htp]
\includegraphics[width=\textwidth]{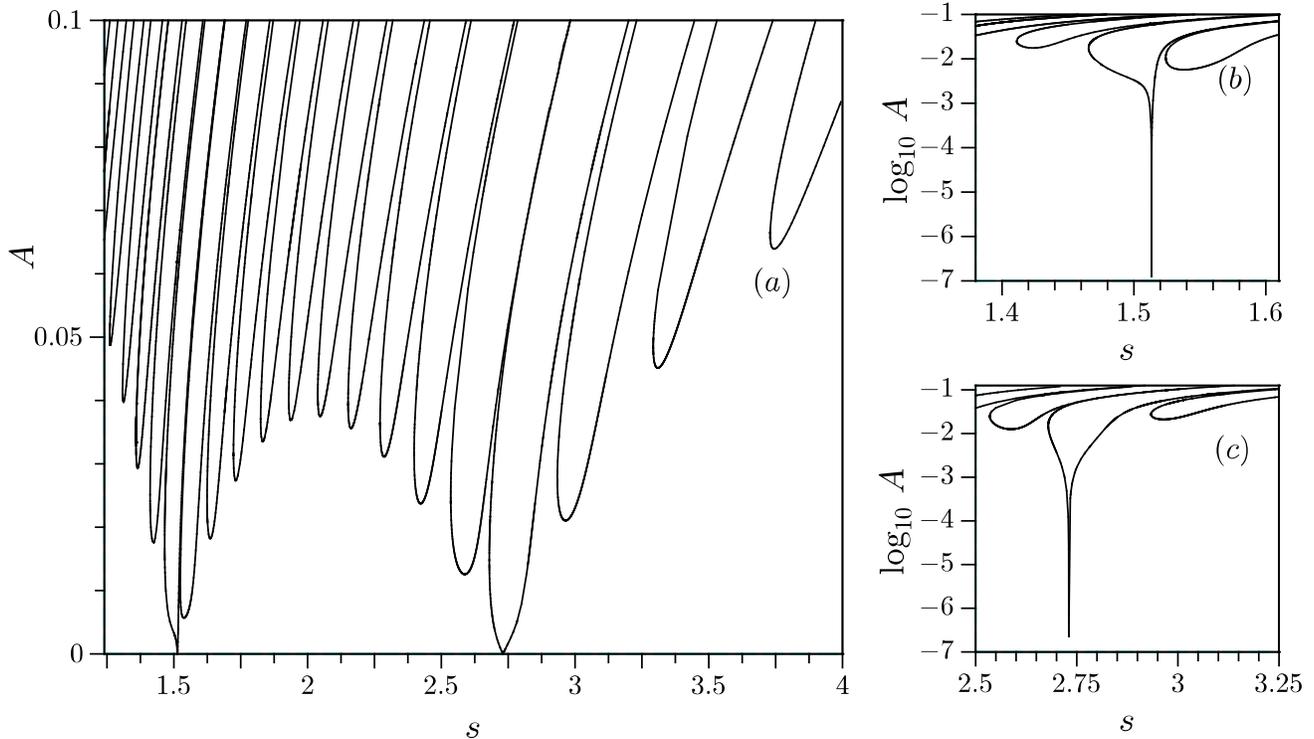}
\caption{{\bf Left}:
Oscillation amplitude $A$ of the nonlocal solution of the nanopteron
with topological charge $Q=-1$ (compression)
as a function or its velocity $s$ and for $L=40$.
 Note that there are some branches that touch zero, but not others.
 Also, for some branches there are  two solutions that correspond
to the same branch. {\bf Right} Zoom of two branches that touch zero
giving rise to a sliding solution or lattice kink. }
\label{fig:curvesAs}
\end{figure*}
%@@@@@@@@@@@@@@@@@@@@@@@@@@@@@@@@@@@@@@@@@@@@@@@@@@@@@@@@@@@@@@@@@@@@@@

We look for the travelling-wave solutions of Eq.~\ref{eq:ddotun} with velocity $s$ given
by $u_n(t)=u(n-st)\equiv u(z)$. For
the unknown function $u(z)$ the set of ordinary ODEs
transform to the following differential-delay equation with the
advanced and retarded terms:
%-----------------------------------------------------------
\begin{eqnarray}
&\mbox{}& s^2 u''(z)=-V'[u(z)]+\nonumber \\&&U'[u(z+1)-u(z)]-U'[u(z)-u(z-1)]~.
\label{eq:dynuz}
\end{eqnarray}
Specific boundary conditions for this model will be discussed later in this section.

Unfortunately, there is no clear analytical technique for
finding solutions of such equations, so that we have to use
numerical methods. A very accurate and simple method for solving
these equations and finding soliton solutions has been
developed in \cite{hmb89pd,eilbeck-flesh1990,duncan-eilbeck1993}.

We look for the solution $u(z)$ in the form
%-----------------------A0-----------------------------------
\begin{equation}
u(z)\simeq u^{(0)}(z) +\sum_{m=1}^k c_m \sin \left (\frac{2\pi m z}{L} \right )~,
\label{eq:uznanop}
\end{equation}
%------------------------------------------------------------
where the second term is a Fourier series of plane waves with
wavenumbers $q_m=2\pi m/L$ and represents any periodic
function with period $L$ in $z$ with coefficients $c_m$ that have to be determined.
The period $L$ is chosen very large so that
the influence of boundaries would be negligible. The integer
part of $L$ will effectively play the role of the chain length
in the original system Eq.~(\ref{eq:ddotun}).  The first term  $u^{(0)}(z)$ is an
initial approximation which is used in order to enhance
the initial guess and/or to enforce the appropriate boundary
conditions.

For this function we specify the boundary conditions as follows
%-----------------------------------------------------------
\begin{equation}\label{eq:boundary}
u(-\infty)=a_1, \quad u(\infty)=a_2,\quad a_1,a_2 \in \Z.
\end{equation}
%-----------------------------------------------------------
The topological charge is defined as $Q=a_2-a_1$, where $Q=-1$ is called an antikink
and  corresponds to a shift in position of a lattice site after the kink has
passed, i.e., to a moving interstitial, while
$Q=1$  corresponds to the opposite phenomena, i.e., a moving
vacancy. For simplicity we will use the term kink for both entities
clarifying the topological charge when necessary.
%############################figure3########################################
%####################################################################
%@@@@@@@@@@@@@@@@@@@@@@@@@@@@@@@@@@@@@@@@@@@@@@@@@@@@@@@@@@@@@@@@@@@@@@
%
%-------------------Figure3 ------------------------------------------
\begin{figure*}[t]
\begin{center}
\includegraphics[width=0.9\textwidth]{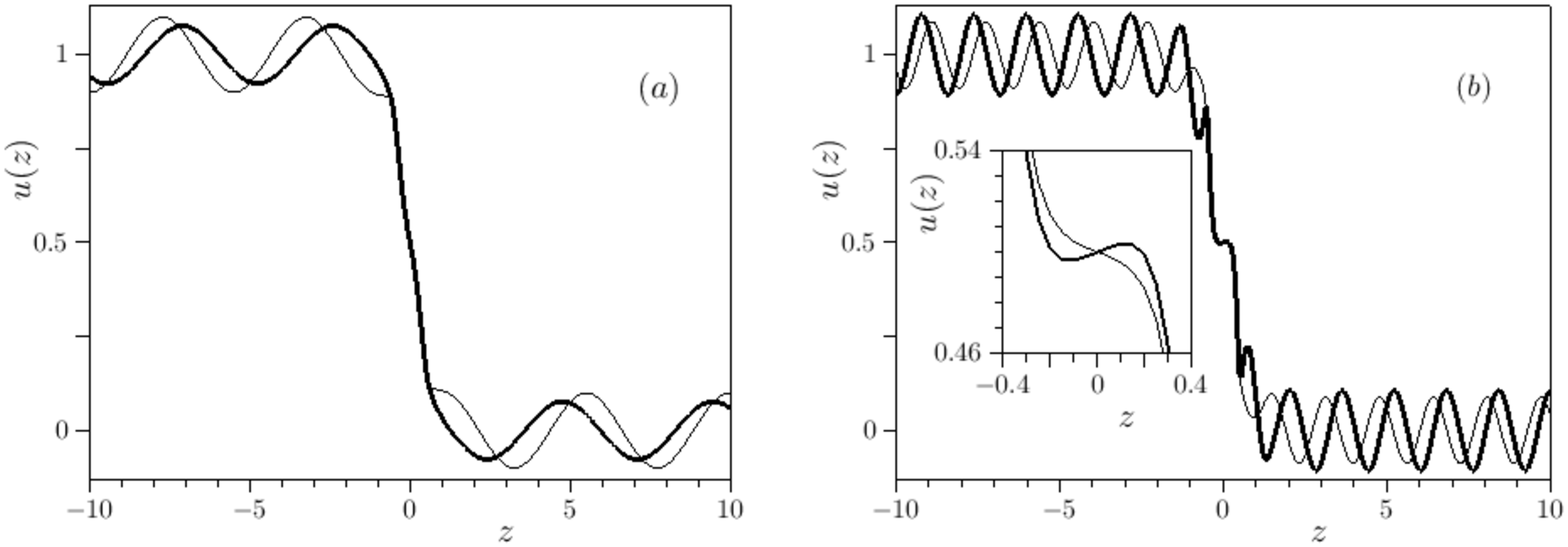}
\end{center}
\caption{
Profiles of the non-local moving solitons (nanopterons) that belong to the
same pair of branches in Fig.~\ref{fig:curvesAs} (thick line: upper branch,
thin line: lower
branch) and have the same velocity $s=3.9$ (a) and $s=1.5$ (b). The inset shows the details
in their core. }
\label{fig:profile_nanopteron}
\end{figure*}
%@@@@@@@@@@@@@@@@@@@@@@@@@@@@@@@@@@@@@@@@@@@@@@@@@@@@@@@@@@@@@@@@@@@@@@
%@@@@@@@@@@@@@@@@@@@@@@@@@@@@@@@@@@@@@@@@@@@@@@@@@@@@@@@@@@@@@@@@@@@@@@
%

In order to extend the method to multi-soliton solutions\,\cite{sze00pd} the initial
approximation $u^{(0)}(z)$ for a general $P$-soliton solution is chosen
in the form
%-----------------------------------------------------------
\begin{eqnarray}
u^{(0)}(z)&=&\frac{a_1+a_2}{2}+\nonumber\\&& \frac{a_2-a_1}{2P}
\sum_{l} \tanh \left [\mu \left (z+l\frac{\delta}{2} \right ) \right ]~,
\label{eq:uztanh}
\end{eqnarray}
\noindent where $l=0, \pm 2, \ldots, \pm (P-1)$ if $P$ is odd and
$l=\pm 1, \pm 3, \ldots, \pm (P-1)$ if $P$ is even.
Here $\mu$ and $\delta$ are trial parameters which will
be determined below for each iteration step by using
various minimization techniques.
The parameter $\mu$ describes the inverse width of one kink
while the parameter $\delta$ can be considered as the
distance between the adjacent kinks.

We substitute the ansatz Eq.~(\ref{eq:uznanop}) into the
original differential-delay-advance equation Eq.~(\ref{eq:dynuz}) and divide the
interval $[-L/2,~L/2]$ into $k$ collocation
points (due to Eq.~(\ref{eq:uznanop})). The solutions we look for
are antisymmetric,
so we can restrict ourselves to the interval $[0,L/2]$)
As a result
we obtain a system of $k$ nonlinear algebraic equations on
unknown coefficients $c_m$ which can be solved by various modifications
of the Newton method. After the finding the coefficients
$c_m$ we reconstruct the shape of the function $u(z)$.

\section{Obtention of travelling nanopterons and kinks with $Q=-1$}
\label{sec:results}

In this section we apply the pseudospectral method defined above
 to the differential-delay equation Eq.~(\ref{eq:dynuz})
and discuss the main properties of its solutions.
We investigate the simplest solution, namely the compression kink
(strictly antikink)  with
$Q=-1$.
To monitor the amplitude of the oscillating tails of the nanopteron
\,\cite{boyd1990-non}  we calculate the tail amplitude
defined by
\begin{equation}
A=\max_{z \in [L/2-\xi,{L/2}]} |u(z)-a_2|
\label{eq:Adef}
\end{equation}
%--------------------------------------------------------------------
which  measures  the amplitude of the tail oscillations in the right end
interval $\xi$. The length of this interval was taken to be  $\xi=L/8$.
For non-oscillating (monotonic) solutions the function Eq.~(\ref{eq:Adef})
 should
be very close to zero and in the limit $L \rightarrow \infty$
this function tends to zero.
The tails of the kink decay quite slowly after the slope and
in order to be sure that $u(z)$ is  horizontal enough
we choose $L$ to be approximately $ 10 u'(0) |a_2-a_1|$.
We find numerically that for the non-oscillating kink solution
 $\varepsilon$ will be almost
undistinguishable from the machine zero.
Minimizing it for each value of system parameters, we
find an appropriate value of the velocity $s$ and the corresponding
monotonic profile while the tail amplitude function Eq.~(\ref{eq:Adef})
reaches a minimum which is almost equal to zero.
%-------------------Figure4 ------------------------------------------
\begin{figure}[t]
\begin{center}
\includegraphics[width=\columnwidth]{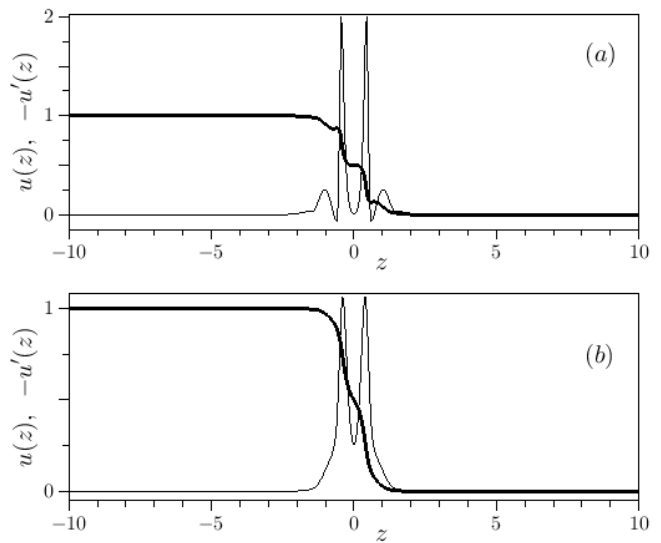}%0.6textwidht
\end{center}
\caption{
Profiles of the sliding kink solutions with $Q=-1$ (compression). ($u(z)$ thick line, compression -$u'(z)$ thin line)
for $s=1.51338$ (a) and $s=2.73025$ (b). The length of the chain is $L=40$. Note that the number of  activated bonds, i.e. the length of
the kink is approximately 4 at (a) and 3 at (b), the two largest velocities. }
\label{fig:profile_kink}
\end{figure}
%@@@@@@@@@@@@@@@@@@@@@@@@@@@@@@@@@@@@@@@@@@@@@@@@@@@@@@@@@@@@@@@@@@@@@@

\subsection{Description of solutions: profiles, amplitudes, velocities and energies}
\label{ssec:descriptionQ_1}
The typical solution of Eq.~(\ref{eq:dynuz}) appears to be a
non-local solution, a bound state of a kink with the plane wave that
moves with the same velocity $s$, also known as a nanopteron\,\cite{boyd1990-non}.
For the fixed value of $L$ we trace the nanopteron solutions
 and monitor the amplitude in their tails $A(s)$ defined in Eq.\,(\ref{eq:Adef}).
 This dependence $A(s)$,
which can be seen as a equation of state,  is
shown in Fig.~\ref{fig:curvesAs} and appears to be a set of pairs of branches
that come from large values of $A$ and that join together at some local minima.
It appears that by gradual increasing $s$
each branch can be continued to an arbitrarily large value of $A$.

The branches amplitude-velocity in each pair as seen in Fig.~\ref{fig:curvesAs} can be describes as an upper one and a lower one.
Each of these branches is a single-valued function and these branches join
    each other at certain bifurcation points where $\textrm{d}A/\textrm{d}s=\infty$.
Alternatively, we can consider the dependence Fig. \ref{fig:curvesAs} as
a function $s=s(A)$. In that case, the pairs of these branches can be
classified as two single-valued functions of $A$ that join together at the
bifurcation point  $\textrm{d}s/\textrm{d}A=\infty$. In this case we should call them
as faster and slower branches since for the fixed amplitude their velocities differ.
Because the kink velocity is the variable parameter in our problem, we choose the
former classification.
%Yaroslav MAYBE TO MOVE THIS PARAGRAPH TO SEC VIA?.
%JuanJun29: moved

Figure~\ref{fig:curvesAs} shows that for a fixed value of the kink velocity $s$
 there exists a countable set of solutions as $A$ is
increased. While $s$ decreases, the branches are packed more densely and
each individual branch $A(s)$ becomes more and more steep.

If the velocity $s$ is fixed, then two different solutions that belong to
one pair of curves differ by the amplitude and the wavelength
of the oscillating tail. Depending on the pair of branches they may
 also differ by the behaviour near the solutions
center. The typical nanopteron profiles are given in Fig.~\ref{fig:profile_nanopteron},
and is can be observed that if a solution lies
on the upper curve of the pair, we may have $u'(0)<0$,
otherwise $u'(0)>0$, as shown in the panel (b). However, for
larger velocities the behaviour in the core remains very similar
with both profiles having $u'(0)>0$.

The wavelength of the nanopteron tail approximately equals to
$2\pi/q_*$, where the wavenumber $q_*$ is the root of the
equation $sq_*=\omega(q_*)$ and $\omega(q)$ is the dispersion
law given in Eq.~(\ref{eq:phonondispersion}). For the two nanopterons
with $s=3.9$ given in Fig. \ref{fig:profile_nanopteron} the
respective wavelength equal $4.47$ for the one marked by the thin
line and $4.70$ for the one marked by the thick line. The solution
of the equation $sq_*=\omega(q_*)$ yields $2\pi/q_*=5.15$.
Close to the sliding velocity the correspondence is much better.
Indeed, for the nanopteron from the lowest branch
of Fig. \ref{fig:curvesAs} at $s=2.75$ we obtain the wavelength
$3.45$ and the root of $sq_*=\omega(q_*)$ gives the same value
up to the third digit. We should keep in mind that the asymptotics
of the nanopteron tails are not exactly phonons, but rather
nonlinear periodic waves. Certainly, for smaller amplitudes they become closer to phonons.

The $A(s)$ behaviour for this lattice is different from the
analogous figure for the discrete Klein-Gordon lattice with
harmonic interactions \cite{zolotaryuk2018springer}, where the
dependence $A(s)$ is generally single valued.

It appears that within the range of velocities $0 < s <4$, the
dependence $A(s)$ hits the $A=0$ axis at several points, namely
$s_1=2.73025$, $s_2=1.51338$, $s_3=0.82501$, $s_4=0.67820$,
$s_5=0.42719$, $\ldots$. As far as velocities $s>4$ are concerned, we
did not detect any zeroes of the $A(s)$ dependence up to $s=12$.
We did not continue into smaller velocity
values because of the bad convergence of the numerical scheme, which is very likely
due to entering into the phonon group velocities zone.

Examples of non-oscillating solitons for two values of the sliding
velocity are given in Fig.~\ref{fig:profile_kink}. Both solutions can be treated as
a bound state of two acoustic solitons, especially well this can be seen from
the $u'(z)$ profiles. The difference between these
solutions is the behaviour in the centre ($z=0$). As $s$ decreases,
the separation between the global minima of $u'(z)$ becomes more
significant. This remains to be true for the rest of the non-oscillating
solutions, those with $s<1$.

If the length $L$ is changed the dependence $A(s)$ also changes, but
insignificantly. The sliding velocities (those at which $A(s)=0$) do not
change. This is logical, because when $A(s)=0$ the boundary conditions become irrelevant.
It has not been possible to find other non-oscillating bound states in spite of
trying different $\delta$ in the initial ansatz Eq.~(\ref{eq:uztanh}),
with one, three or other number of kinks in one bound state.

\subsection{Energies}

We set the zero of the energy at the equilibrium distances, that is, the energy is given by
$E=H-U_0$,
where $U_0$ is  the sum of the potential interactions at the equilibrium distances, $U_0=[\sum_n U(u_n-u_{n-1})]_{u=0}$. Note that the on-site
potential $V(u_n)$ is already zero for $u_n=0$.

The values of scaled and physical velocities are given in Tab.~\ref{tab1}.
\begin{table}[htb]
\begin{center}
\begin{tabular}{|l||c|c|c|c|c|}
\hline
Notation & $s_5$ & $s_4$ & $s_3$ & $s_2$ & $s_1$ \\
\hline
{\it V} (scaled) & 0.4271903 & 0.678198 & 0.825005 &  1.513382 & 2.73025 \\
\hline
{\it E} (scaled)   &  9.177956  &  9.230015 & 9.241154 &  9.10141 & 9.427962\\
\hline
{\it V} (km/s) & 1.1171&    1.7734 &   2.1573 &   3.9574 &   7.1394\\
\hline
{\it E} (eV) &  25.4648  & 25.6093  & 25.6402  & 25.2525  & 26.1585\\
\hline
\end{tabular}
\end{center}
\caption{Scaled and physical velocities and energies  of the localized kinks with $Q=-1$. Note that in spite of the different velocities, their energies
are very similar}
\label{tab1}
\end{table}
Note that in spite of the different velocities of the sliding solitons,
their energies appear to vary by less than $1\%$. This is logical because the ions need to have the same energy to overcome the same potential barrier and they need the same energy increase to be two of them fairly close in the same potential well.

%%@@@@@@@@@@@@@@@@@@@@@@@@@@@@@@@@@@@@@@@@@@@@@@@@@@@@@@@@@@@@@@@@@@@@@@
%%@@@@@@@@@@@@@@@@@@@@@@@@@@@@@@@@@@@@@@@@@@@@@@@@@@@@@@@@@@@@@@@@@@@@@@

\section{Stability of kinks and nanopterons with $Q=-1$}
\label{sec:stabilityQ_1}
%@@@@@@@@@@@@@@@@@@figure5 @@@@@@@@@@@@@@@@@@@@@@@@@@@@@@@@@@@@@
\begin{figure*}[htb]
\begin{center}
\includegraphics[width=0.48\textwidth]{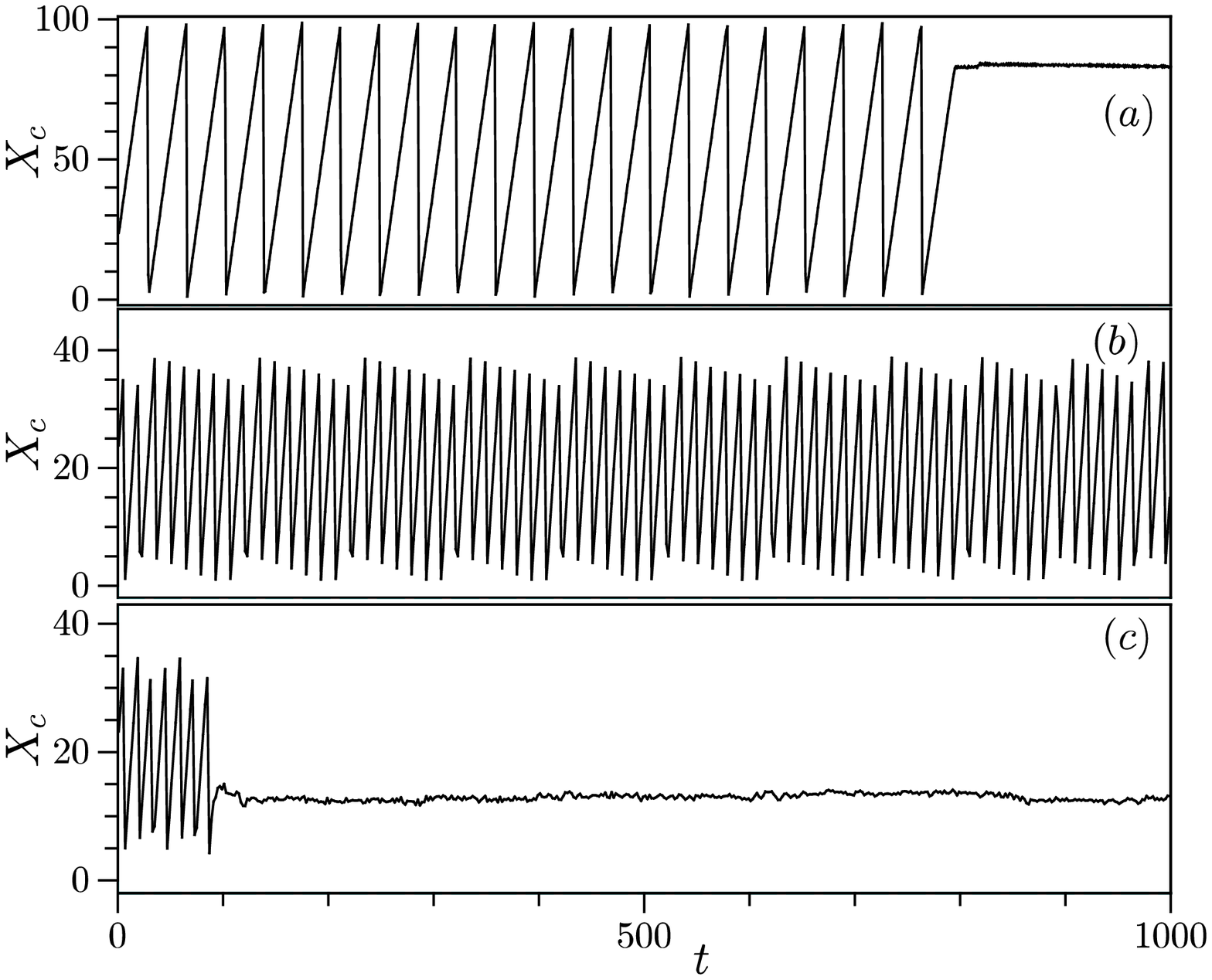}%[width=12.cm,angle=0]
\includegraphics[width=0.48\textwidth]{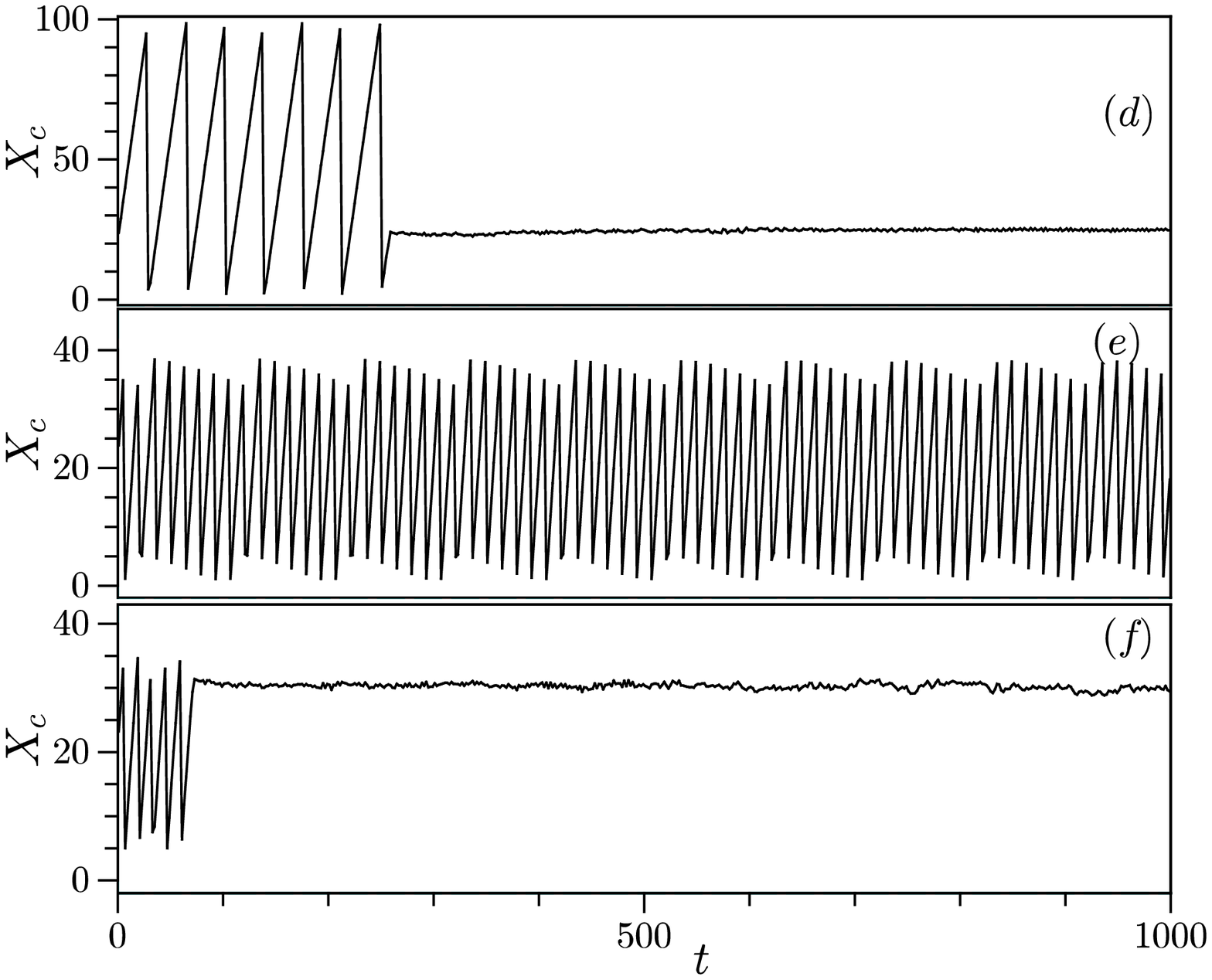}
\end{center}
\caption{Center of mass (\ref{eq:cmass}) evolution for:
(a) the kink with the
sliding velocity $s=2.73025$ with the randomly perturbed initial velocities
according to Eq.~(\ref{eq:rand2a})
and
$\epsilon\simeq 0.01$ of the order of thermal perturbation at room temperature; (b)
Nanopterons with $\epsilon=0.001$ in Eq.~(\ref{eq:rand2a}),
$s=2.8$ and $A=0.0079$; (c) Nanopteron with $s=3$ and $A=0.064$.
Panels (d)-(f) correspond to the same as (a)-(c) but with the
initially perturbed displacement field as in Eq.~(\ref{eq:rand2b}).
}
\label{fig:kink_random_stability}
\end{figure*}
%@@@@@@@@@@@@@@@@@@@@@@@@@@@@@@@@@@@@@@@@@@@@@@@@@@@@@@@@@@@@@@@@@@@@@@
%   yaroslav{In this figure I have used $\epsilon=0.01$, see the old version.}

\begin{figure*}[p]
\begin{center}
\includegraphics[width=1.0\textwidth]{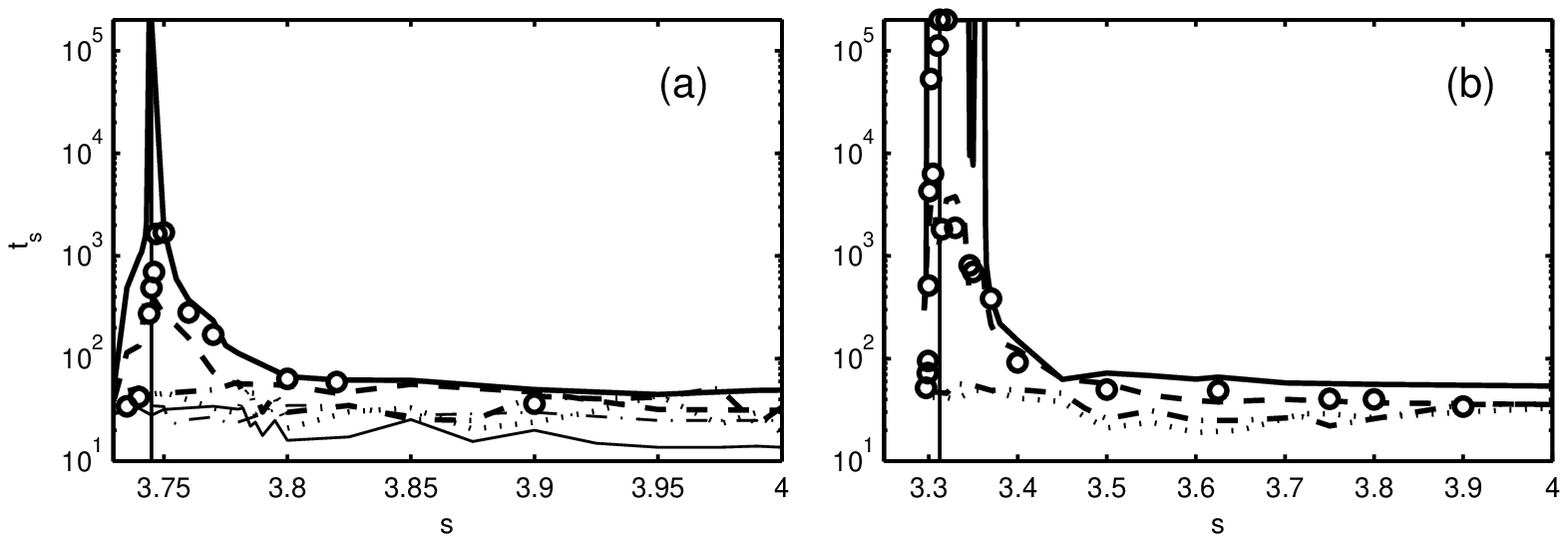}\\
\includegraphics[width=1.0\textwidth]{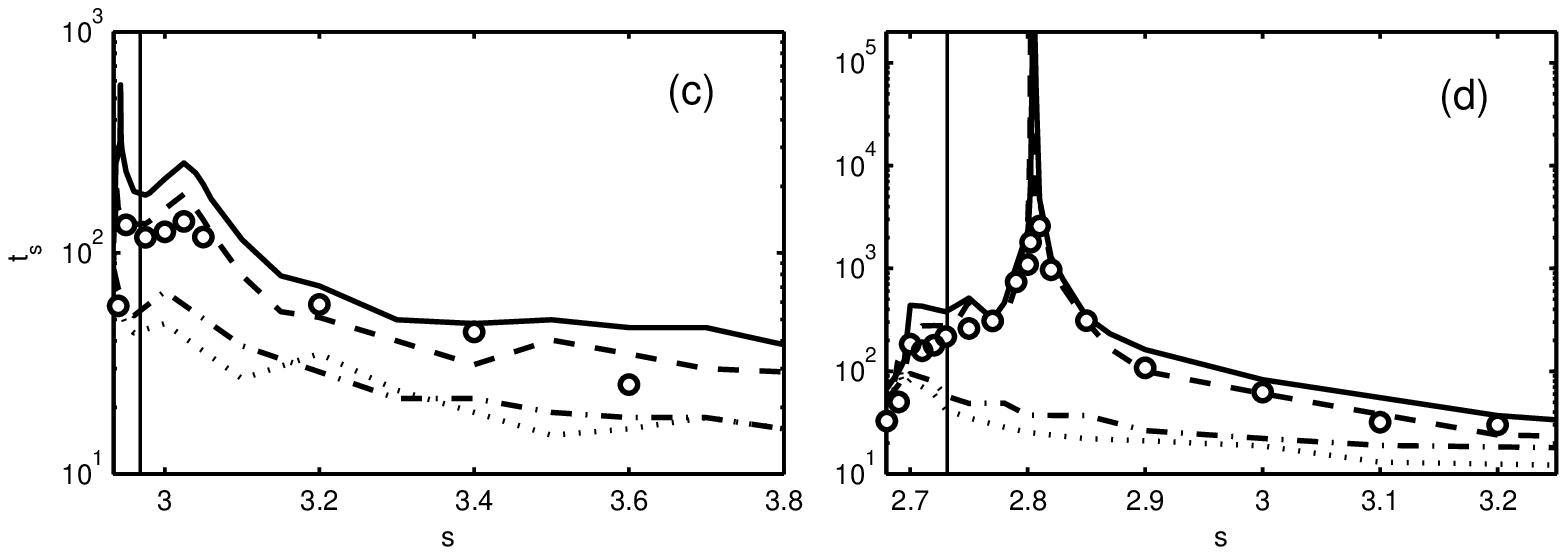}\\
\includegraphics[width=1.0\textwidth]{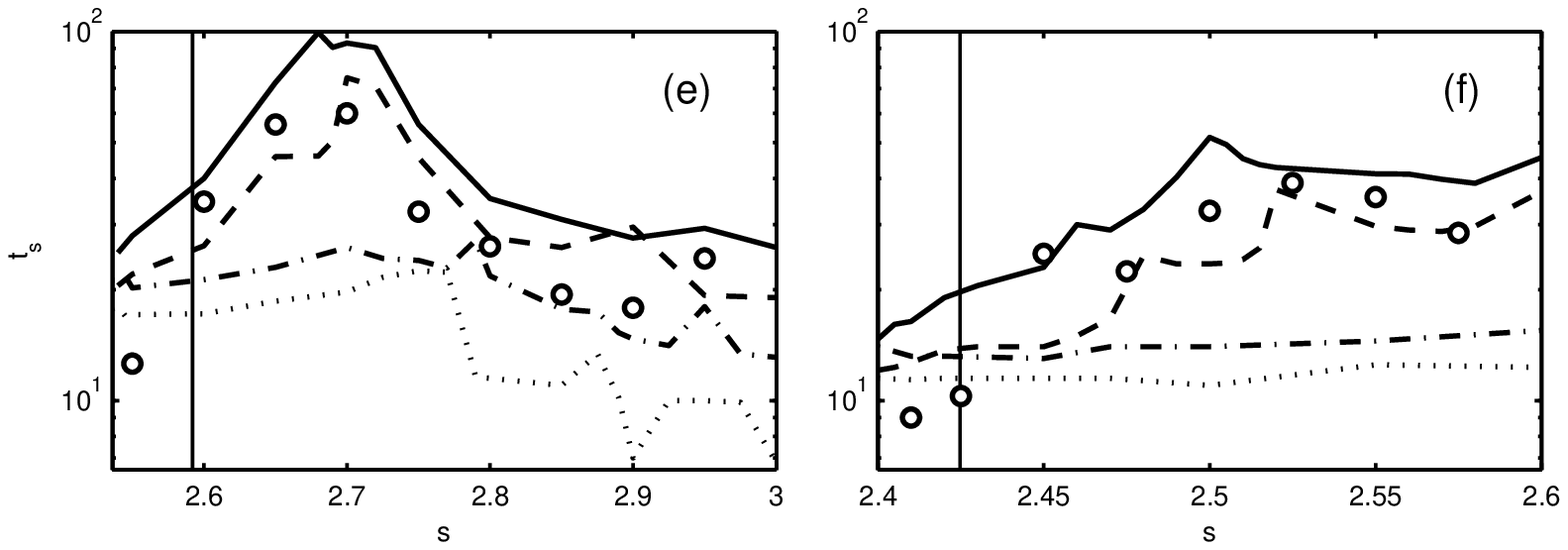}\\
\caption{Survival time  $t_s$ (see text) as a function of the velocity for
different branches of the $A(s)$ dependence represented in Fig. \ref{fig:curvesAs}.
The panels (a)-(f) correspond to the pairs of curves 1-6 starting from the right, i.e., from higher velocities.
The lower curve is generally more stable than the upper one for the same branch.
Lower branches: thick continuous line for the perturbation parameter $\epsilon=0.001$, dashed for $\epsilon=0.01$;
upper branches: dash-dotted line for $\epsilon=0.001$, dotted line for  $\epsilon=0.01$.
Also for panel (a) there are two thinner lines at the bottom for $\epsilon=0.1$, continuous for
the upper branch and dotted for the lower branch.
Circles ($\circ$) mark results for the randomly perturbed chain described in Eq.~(\ref{eq:rand2b})
with $\epsilon=0.01$.
Vertical thinner lines mark the
position of the local minimum of the respective $A(s)$ dependence.
%Red and black colors correspond to upper and lower curves, respectively.
%The thickest lines
%correspond to $\epsilon=0.1$ [present only in panel (a)], the intermediate
%ones to $\epsilon=0.01$ and the thinnest to $\epsilon=0.001$.
}
\label{fig:global_stability}
\end{center}
\end{figure*}

Stability of the moving sliding kinks with $Q=-1$ has been checked by integrating numerically
the dynamical equations of motion where the exact or perturbed solution
of Eq.~(\ref{eq:dynuz}) was used as an initial condition.
The perturbation was introduced by distorting the velocity field
${\dot u}_n(0)$ in the several ways, for example
\begin{equation}\label{eq:regular}
{\dot u}_n(0) \to (1+\epsilon) {\dot u}_n(0)~,
\end{equation}
where $\epsilon$ is the perturbation parameter. The displacement field $u_n(0)$ was kept unchanged.
Alternatively, the displacement field $u_n$ can be excited in the
similar manner while the velocity field was undistorted.
%juan-jun29
Although, the two methods seem quite different they lead to similar results for  the stability as can be seen in Fig.~\ref{fig:global_stability}.
The perturbation parameter has been varied from -0.1 to 0.1. The boundary conditions were
periodic. Only the kink with fastest velocity $s_1=2.73025$ is stable for positive values of $\epsilon=0.1$ but
gets pinned for negative $\epsilon=-0.01$. This is due to the fact that being the perturbation proportional to the particle velocities
in the kink core, which are large, the diminution of the velocities takes too much energy away from the kink.

A different perturbation can be done with a random variation of the
initial displacements or velocities:
%-----------------------------------------------------------------------
\begin{eqnarray}\label{eq:rand2a}
&&{u}_n(0) \to \epsilon(\xi_n-1/2)  + {u}_n(0),
~~~n=1,2,\ldots,N\,,\\
&&{\dot u}_n(0) \to \epsilon(\xi_n-1/2)  + {\dot u}_n(0),
~~~n=1,2,\ldots,N\,, \label{eq:rand2b}
\end{eqnarray}
%-----------------------------------------------------------------------
where $\{\xi_n\}_{n=1}^{N_l}\in [0,1]$ are random numbers. This perturbation is of the order of the thermal velocities
at room temperature for $\epsilon\simeq 0.1$.
The kink also shows significant robustness. The center of mass
\begin{equation}\label{eq:cmass}
X_c(t)=\frac{\sum_{n=1}^N {n\cal E}_n}{\sum_{n=1}^N {\cal E}_n},
\end{equation}
dynamics is given in Fig.~\ref{fig:kink_random_stability}a. Here the
energy density ${\cal E}_n$ is defined in Eq.~(\ref{eq:Htotal}).

The stability of nanopterons depends on many details such
as speed, amplitude and also the lattice size.
Generically speaking, nanopterons with fast velocities $s\gtrsim2.5-3$ can travel 10-25 times the whole lattice before getting pinned, while slower nanopterons $s\lesssim 2$ travel
much shorter distances or get pinned in the lattice from the very beginning.
If the nanopteron speed is close to the sliding velocity, it can
be significantly robust, travelling successfully around the lattice with $N=40$
more than 60 times (see Fig. \ref{fig:kink_random_stability}b).
Another nanopteron solution with the velocity $s=3$ and same perturbed
initial conditions has managed to travel much shorter distance as shown
in Fig. \ref{fig:kink_random_stability}c. In all three cases of Fig. \ref{fig:kink_random_stability} the periodic boundary conditions
$u_{n+N}=u_{n}+Q$ were used.
%The results depend, of course, of the many details as amplitude and also the lattice size.
From the physical point of view, we can conclude that  nanopterons can play a role as a transient state before the sliding kink velocity is established. It is, of course, impossible, to produce an exact nanopteron in a physical process as it should include a very large part of the lattice, but approximations to them might happen.

%yaroslav-begin
To get better understanding of the nanopteron and kink stability, we
have performed the global stability studies. For that, we need to follow each of the
branches that appear in Fig.~\ref{fig:curvesAs}. We have used the homogeneous velocity perturbation,
described in Eq. (\ref{eq:regular}).
In this way, we can
 exclude the dependence on the particular realization of disorder.
We also have used the random perturbation scheme (\ref{eq:rand2b})
for one fixed disorder realization.
To characterize the robustness of the solitary wave, we introduce the
{\it survival} time $t_s$ for which the progressive propagation
with the constant velocity is sustained:
%--------------------------------------------------------------------
\begin{equation}
X(t)=\left \{
\begin{array}{ccc}
&& X_0+st, t<t_s ,\\
&& \mbox{pinned}, t>t_s
\end{array}
\right . ~.
\end{equation}
%--------------------------------------------------------------------
For example, in the case shown by Fig. \ref{fig:kink_random_stability}a
$t_s\sim 800$.
The logarithmic dependence of this critical time is given in Fig. \ref{fig:global_stability}.
We numerate the pairs of curves in Fig. \ref{fig:curvesAs}
from right to the left. Thus, the first pair is positioned
in the top right corner
of that figure. The pair number $4$ contains the non-oscillating kink
with $s_1=2.73025$. The second non-oscillating kink belongs to the
pair $15$. Then we take each branch (upper and lower) from each pair,
and compute the time $t_s$ while the velocity is varied.
For the first pair we have used three values of the perturbation
$\epsilon=0.1$, $\epsilon=0.01$ and $\epsilon=0.001$. This case is depicted
in Fig.\ref{fig:global_stability}a. As one can see, for the value $\epsilon=0.1$
the original solution gets destroyed too quickly and not much is seen
on the logarithmic scale.
%@@@@@@@@@@@@@@@@@@@@@@@@@@@@@@@@@@@@@@@@@@@@@@@@@@@@@@@@@@@@@@@@@@@@
%@@@@@@@@@@@@@@@@@@@@@@@@@@@@@@@@@@@@@@@@@@@@@@@@@@@@@@@@@@@@@@@@@@@@
Therefore we drop it for the next cases. The main results of the
global stability studies are the following:
\begin{itemize}
\item Only the nanopterons from the lower
branches can be truly stable,
the survival time for the upper branch solutions never exceeds $10^2$
 That means it can hardly travel more than $100$ lattice sites. On the contrary,
the lower branch solutions can be extremely robust with $t_s> 2 \times 10^5$
(the computation was stopped at those times).
\item The intervals of high nanopteron stability are positioned at the
lower parts of the $A(s)$ dependencies (for small amplitudes).
\item High robustness of the nanopterons is seen for the high-speed
branches, especially for the branches 1,2 and 4.
The branch 4 contains
 a non-radiating kink, however the most stable solutions are nanopterons
with velocities $s\in [2.805,2.807]$ that slightly exceed the non-radiating
kink velocity $s_1$.
%juanJun29
This can be understood easily as the faster nanopteron can slow down because of disorder keeping its stability.
The intervals of the extremely high kink robustness are very close
to the respective minima of the $A(s)$ curve for the pairs 1 and 2.
Thus, the most stable nanopterons have the smallest tail amplitude.
However, as commented above, this is not true for the branches
that contain the sliding velocity, where nanopterons with $A>0$ are more stable.

\item Nanopterons starting from the branch $5$ and further on appear to
be much less robust (see Figs. \ref{fig:global_stability}e-f). We did not
plot the $t_s(s)$ dependencies for the rest of the branches. As we move
towards lower velocities, the dependencies look qualitatively the same as
in \ref{fig:global_stability}e-f. The values of $t_s$ further decrease
to the values $t_s \sim 10$ or even less. At some point (starting from the pair
number $12$) it becomes
impossible to define $t_s$ as nanopterons can travel only several sites,  which confirms that nanopterons and kinks
can only be stable when their velocity is larger than the minimum phonon velocity in the
first Brillouin zone $v_{ph,min}=1.68$.
\end{itemize}
We also note that random perturbations of the initial conditions (\ref{eq:rand2b})
give qualitatively the same result as the homogeneous perturbations. The survival times are plotted as circles in Fig.~\ref{fig:global_stability}.

%juan jul1 figure suppressed@@@@@@@@@@@@@@@@@@@@@@@@@@@@@@@@@@@@@@@@@@@@@@@@@@@@@@@@@@@@@@@@@@@@
%\begin{figure}[htb]
%\begin{center}
%\includegraphics[width=1.0\columnwidth]{fig7.eps}
%\caption{
%{\bf Top: } Survival time $t_s$
%as a function of the soliton velocity for
%lattice size $L=40$ (thick lines) and $L=50$ (thin lines).
%Black curves show the survival time for the bottom branch that
%has a non-zero velocity (black lines),
% the nearest left-side (red) and right-side branch (blue).
%{\bf Bottom:} Nanopteron amplitude as a function
%of velocity for $L=40$ (thick lines) and $L=50$ (thin lines).
%The perturbation parameter was $\epsilon=0.01$ everywhere.}
%\label{fig:global_stability2}
%\end{center}
%\end{figure}
%@@@@@@@@@@@@@@@@@@@@@@@@@@@@@@@@@@@@@@@@@@@@@@@@@@@@@@@@@@@@@@@@@@@@

%yaroslav
The change of the lattice size does not bring any principal
changes to the stability picture.
%juan jul1: commented below and also other placed due to suppression of figure global_stability2
%This can be seen from Fig. \ref{fig:global_stability2}a,
%where the survival time is plotted as a function of the nanopteron
%speed for different values of the lattice size. The method
%of the $t_s$ computation was the same as for the previous figure.
As expected there are changes of the nanopteron tail amplitudes $A(s)$ for the same velocities with respect to
the lattice size, which tend to disappear in the vicinity of the sliding kinks.
The nanopterons are more robust for
the lowest part of the $A(s)$ dependencies, while
%We have followed only the lower (bottom) branches of the $A(s)$
%dependencies because, as revealed in Fig. \ref{fig:global_stability},
nanopterons that belong to the top branches are much less stable.
%juan jul1 commented
%The maximum of the
%$t_s(s)$ dependence shifts because the existence interval of the
%particular branch shifts as well (see Fig. \ref{fig:global_stability2}b).
%end yaroslav

A general conclusion is that it seems that both sliding kinks and
nanopterons are only stable at large velocities of the order of the
fastest and only stable crowdion with $s_1=2.73$ and $q=0.584\pi$ which
is close to twice the speed of sound in the lattice without substrate $2c_s$
and above five times larger
than the group velocities. This is also valid for the double crowdions
described in Sect.~\ref{sec:double}. We have not a mathematical proof but the
reason for it seems evident looking at Fig.~\ref{fig:phonons}: kinks with velocities that do not correspond to a wavevector in the first Brillouin zone are not stable and get pinned to the lattice because although they are mathematical solutions they are an artificial one.

The wavenumber associated with the tail of both the crowdion and bi-crowdion do not coincide although it is not very far from the {\em magic} wave number $q=2\pi/3$\,\cite{kosevich04} which shows very good propagation in FPU lattices because it is an exact solution of the FPU lattice when the interaction potential can be approximated to a polynomial of degree four\,\cite{kosevich93}. This variation is certainly due to the existence of the on-site potential and the peculiar, but realistic, characteristics of the model with hard on-site potential and two repulsive interactions.
%@@@@@@@@@@@@@@@@@@@@@@figure7@@@@@@@@@@@@@@@@@@@@@@@@@@@@@@@@@@@@@@@@@@@@@@@@
%----------------------------------------
\begin{figure}[b]
\begin{center}
\includegraphics[width=\columnwidth]{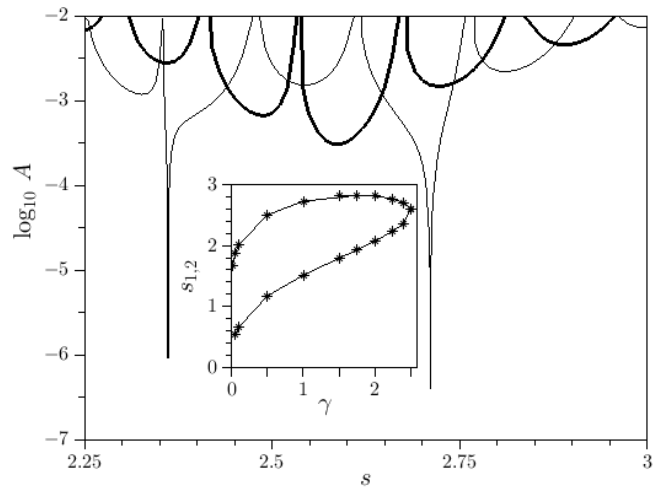}
\end{center}
\caption{Oscillation amplitude $A$ of the nonlocal solution
as a function or its velocity $s$ and for $L=40$. Thin line corresponds
to $\gamma=2.4$ and thick line to $\gamma=2.51$. The inset shows
the dependence of the two largest sliding velocities on the
on-site potential amplitude $\gamma$.
 }
\label{fig:As_gamma_neq1}
\end{figure}
%@@@@@@@@@@@@@@@@@@@@@@@@@@@@@@@@@@@@@@@@@@@@@@@@@@@@@@@@@@@@@@@@@@@@@@
%yaroslav
The global stability results can be explained in the following
way. Stable nanopteron solutions are observed for small tail amplitudes and
only for those branches with $s\gtrsim s_1$, where $s_1$ is the largest
sliding velocity ($A(s_1)=0$). Large tail amplitudes can trigger
the modulational instability in the nanopteron asymptotics that must
eventually destroy the moving solution. Careful analysis of the
$A(s)$ dependence (Fig. \ref{fig:curvesAs}) reveals that the pairs
of branches with $s\gtrsim s_1$ are more isolated from each other.
As $s$ decreases these pairs become more and more densely packed. Thus,
if we take a solution from one of those branches and perturb it,
an even small perturbation can
easily bring us to another solution, with smaller velocity,
and, further, to the next one. If the amplitude $A$ is quite large, the
branches are even more densely packed and this hopping can take place faster.
On the contrary, if the solution from the more isolated branch is perturbed,
one needs a much stronger perturbation to get to the neighbouring solution.
%end yaroslav

%####################################################################
%####################################################################
\section{Variation from the physical parameters}
\label{sec:otherparameters}
%####################################################################
%####################################################################

Although the values of the parameters in the system have been obtained
from physical consideration, it is interesting to know what happens if
they change.

\subsection{Change of the magnitude of the on-site potential}

Suppose $\gamma\neq 1$ in Eq.~(\ref{eq:Vu}), then there is not a
significant change in the global picture, only there are different
values of the
sliding velocities as represented in Fig.~\ref{fig:As_gamma_neq1}, and only one of them is stable, the fastest one. As it can be seen in the inset of
the same figure the stable velocity $s_1$ increases monotonically with $\gamma$, which is probably related with the increase of the phase velocity of the phonons. However, for  a value of $\gamma\simeq2.5$ the two velocities coincide and
they no longer exist for larger $\gamma$.  Therefore, somewhat against intuition, the increase of the magnitude of the on site potential increases the
velocity of the sliding kinks but after some value makes it impossible, restoring again the intuition.

\subsection{Suppression of the short-range ZBL potential}

If the short range ZBL potential is suppressed in Eq.~(\ref{eq:Vu}) there are no localized solution with
velocity $s>1$. The picture for smaller velocities in quite complicated because the branches of the different non-local solutions are very dense.
The main conclusion can be that the short-range potential is an important factor for the existence of kinks with fast velocity.
%YURIY: compare with sine-gordon?
%%%%%%%%%%%%%%%%%%%%%%%%%%%%figure8%%%%%%%%%%%%%%%%%%%%
\begin{figure}[b]
\begin{center}
\includegraphics[width=\columnwidth]{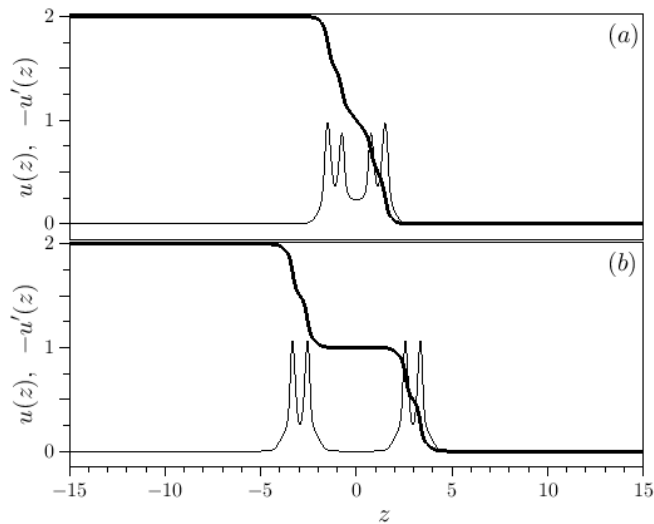}
\end{center}
\caption{
 %yaroslav
 Profile of the stable sliding double crowdion with the largest velocity
and different distances between the kinks. The velocities are $s=3.2619$ (a)
and $s=2.73036$ (b).
 %end yaroslav
 }
\label{fig:bicrowdion}
\end{figure}
%--------------------------------------------------------------------

\section{Double crowdions, $Q=-2$ and other topological charges}
\label{sec:double}
\begin{figure}[b]
\begin{center}
\includegraphics[width=\columnwidth]{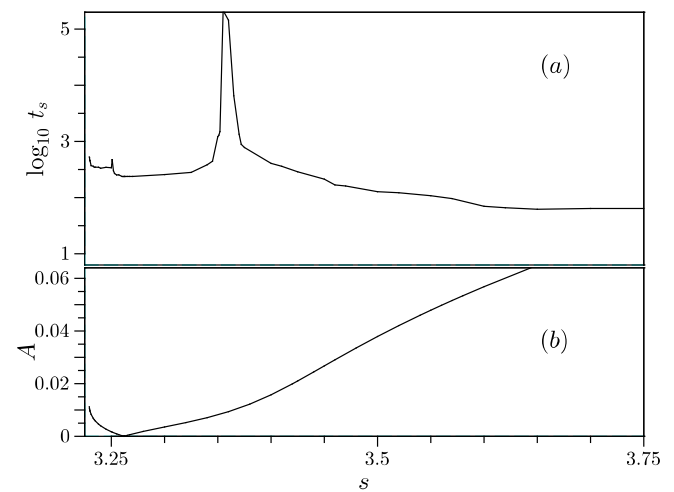}
\end{center}
\caption{{\bf Top}: Survival time as a function of the velocity for the double nanopteron
with $Q=-2$, $L=50$. {\bf Bottom:} Double nanopteron tail amplitude as a function of
velocity.}
\label{fig:bicrowdion_stability}
\end{figure}

We have also studied the case with topological charge $Q=-2$ which corresponds to a
travelling state with two extra particles or a double crowdion.
 %yaroslav
There exist bi-crowdions that consist of two kinks with a distance between them. We
have managed to obtain solutions with different distances as different
velocities, as shown in Fig.~\ref{fig:bicrowdion}. One of them with two
crowdions being quite close and with velocity $s_{D1}=3.2619$ and energy $E\simeq 20$ (56\,eV)
Another one has larger distance approximately six lattice units
between their cores and
 with velocities very similar to the single one with the first sliding velocities
$s_{D2}=2.73036$ and $s_{D3}=1.51339$ very similar to the $Q=-1$ case and
also approximately the double of energy, i.e. $E\simeq18.8$ (52\,eV).
The double crowdions with velocities $s_{D1}$ and $s_{D2}$ are stable while the other is unstable.
In Fig. \ref{fig:bicrowdion_stability} we show the survival time dependence
on the double nanopteron  velocity corresponding to the branch with $s_{D1}$. It has been computed with the same procedure as
for Figs.~\ref{fig:global_stability}
 %juan jul1 commented: and \ref{fig:global_stability2}
 for $\epsilon=0.01$.
This dependence looks very similar to Fig.~\ref{fig:global_stability}d where the
same branch of the $Q=-1$ case has been studied. Both the bi-crowdion
and the double nanopteron solutions can be treated as stable.
%The numerical method has not been able to find sliding double crowdions
%with other separation, which means that they are a single entity and not one
%entity travelling behind the other. Its profile can be seen in Fig.~\ref{fig:bicrowdion}.
%end yaroslav

Note that N-crowdions have been recently found with molecular dynamics in FPU Morse lattices\,\cite{chetverikov2017,dmitrievNsuper2017}. We can speculate that also triple, quadruple,. etc. crowdions also exist, similar to cnoidal waves in KdV equation.
As there are many parameters involved we cannot discard that other bi-crowdions also exist. Also, other N-crowdions, although our search for tri-crowdions ($Q=-3$) has been unsuccessful. On the other hand, the energies of N-crowdions, of the order of N times the energy of a single one is too high to correspond to a realistic lattice excitation.

%%%%%%%%%%%%%%%%%%%%%%%%%%%%figure 9%%%%%%%%%%%%%%%%%%%%

The large energy of a double crowdion can only be provided by one type of $\beta$ decay of $^{40}$K: electron capture followed by conversion electron, where the nucleus decay to an excited state of $^{40}$Ar emitting a 1460\,KeV gamma photon which kicks out a shell electron. The outcome is Ar$^{++}$, with 49.7\,eV, which is close to the double crowdion energy. It is very likely that after the first collision it  de-ionizes due to its high ionization energy and brings about a potassium ion with 57\,eV. This type of decay is only 0.01\% but it is monochromatic, differently to the $\beta^-$ decay where only a fraction of recoils have enough energy to produce nonlinear excitations\,\cite{archilla-kosevich2015b-book,archillaLoM2016}. Of course, this energy can be found in other processes and particularly in the experiments where nonlinear excitations are produced by $\alpha$ bombardment\,\cite{russell-experiment2007,russell-archilla2017}.

%%%%%%%%%%%%%%%%%%%%%%%%%%%%%%%%%%%%%figure10
\begin{figure}[t]
\begin{center}
\includegraphics[width=0.8\columnwidth]{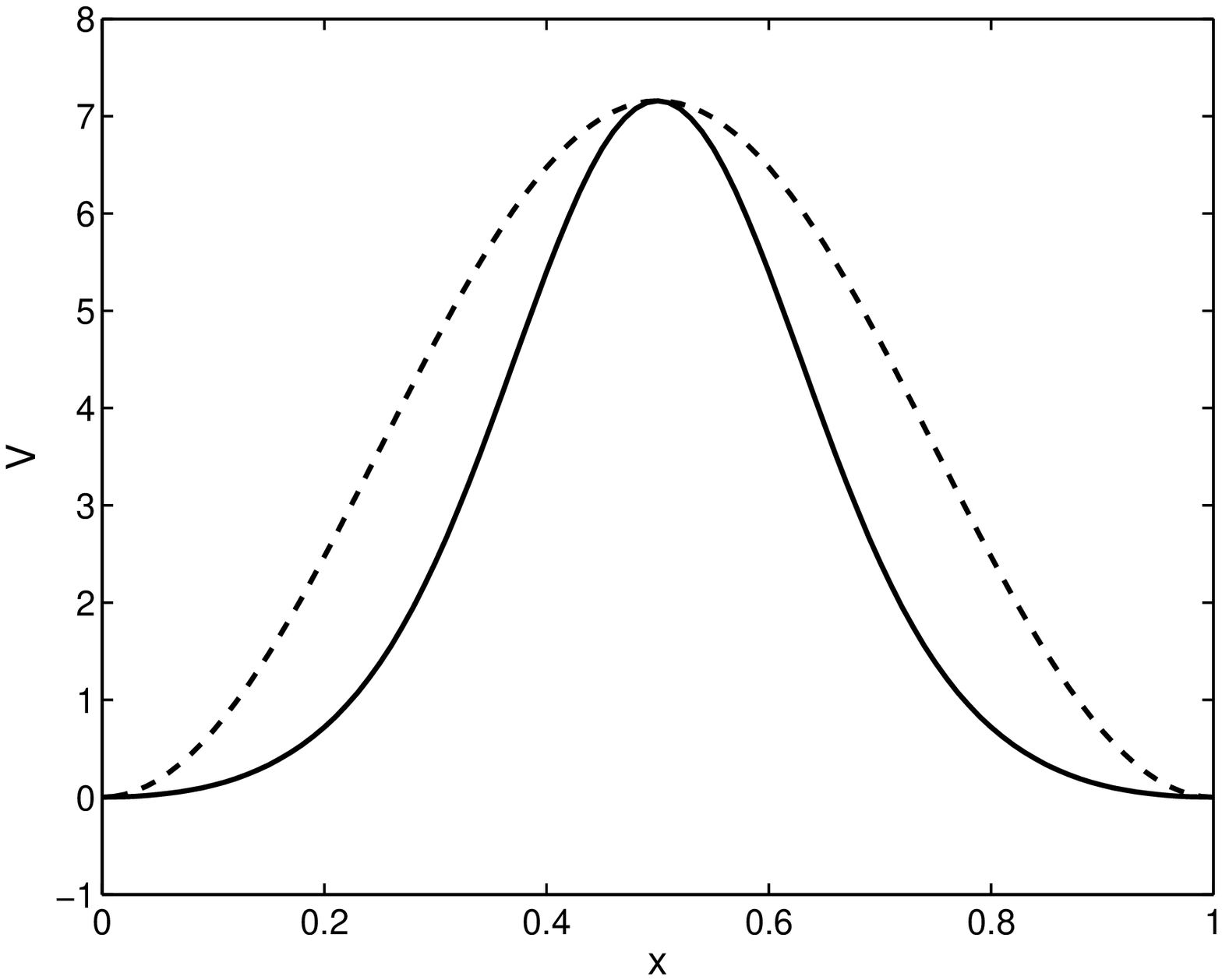}\\
\includegraphics[width=0.8\columnwidth]{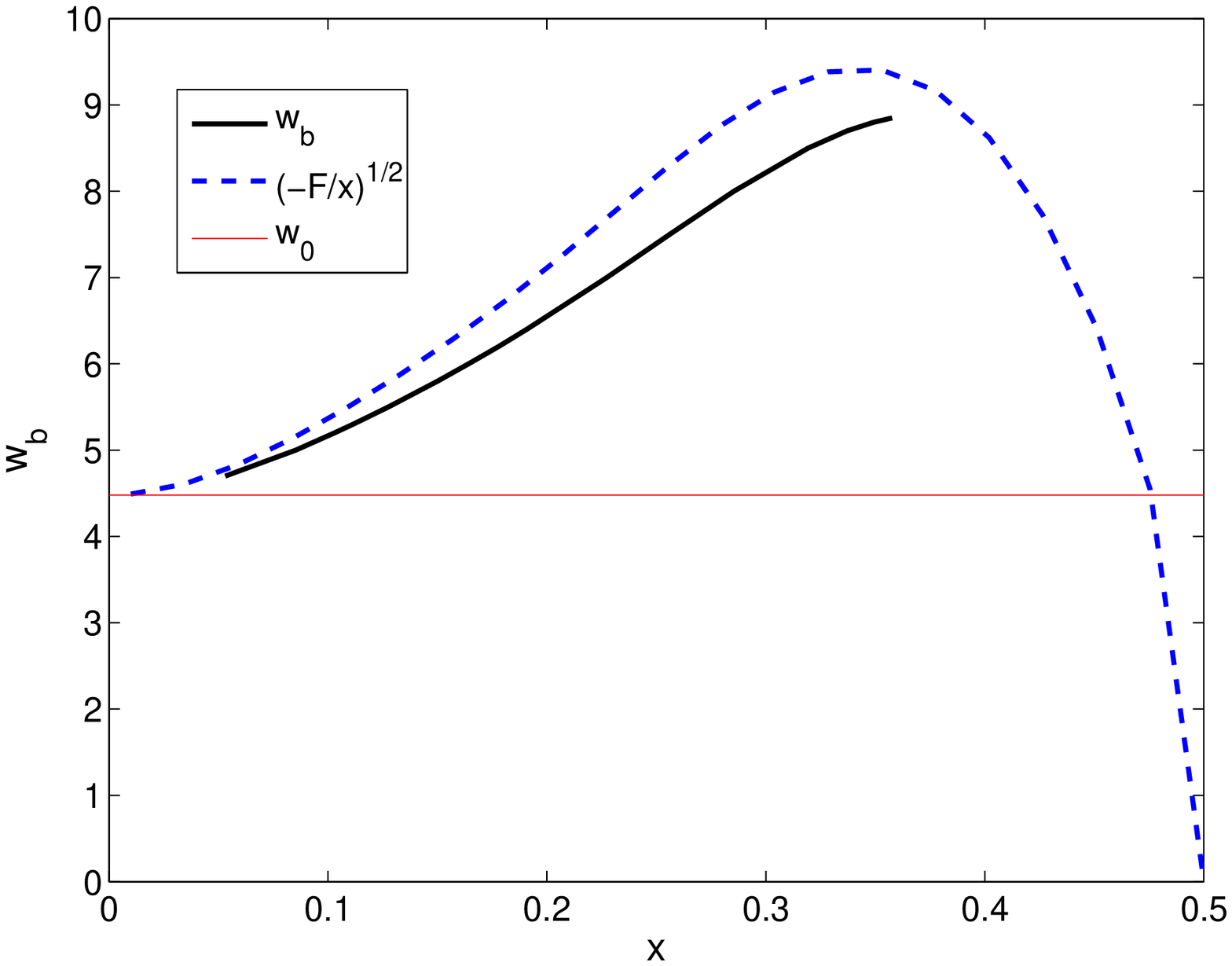}
\end{center}
\caption{{\bf Top}: Profile of the substrate potential (---) compared with the sinusoidal potential $V_\textrm{s}(x)=0.5 V_0(1-\cos(2\pi x))$ ($-\,-$). The substrate potential grows faster that the tangent harmonic potential which makes it {\em hard} for $x\lesssim 0.35$, an essential difference with the sinusoidal one. {\bf Bottom}:   Frequency of the single oscillator as a function of the amplitude $A$ (---) and the square root of the ratio of the restoring force and the displacement ($-\cdot -$)
which would be the frequency for a linear oscillator with that constant elastic constant. The horizontal thin line marks the frequency of the linear oscillator
}
\label{fig:wA_single}
\end{figure}

%%%%%%%%%%%%%%%%%%%%%%%%%%%%%%figure11
\begin{figure}[t]
\begin{center}
\includegraphics[width=0.8\columnwidth]{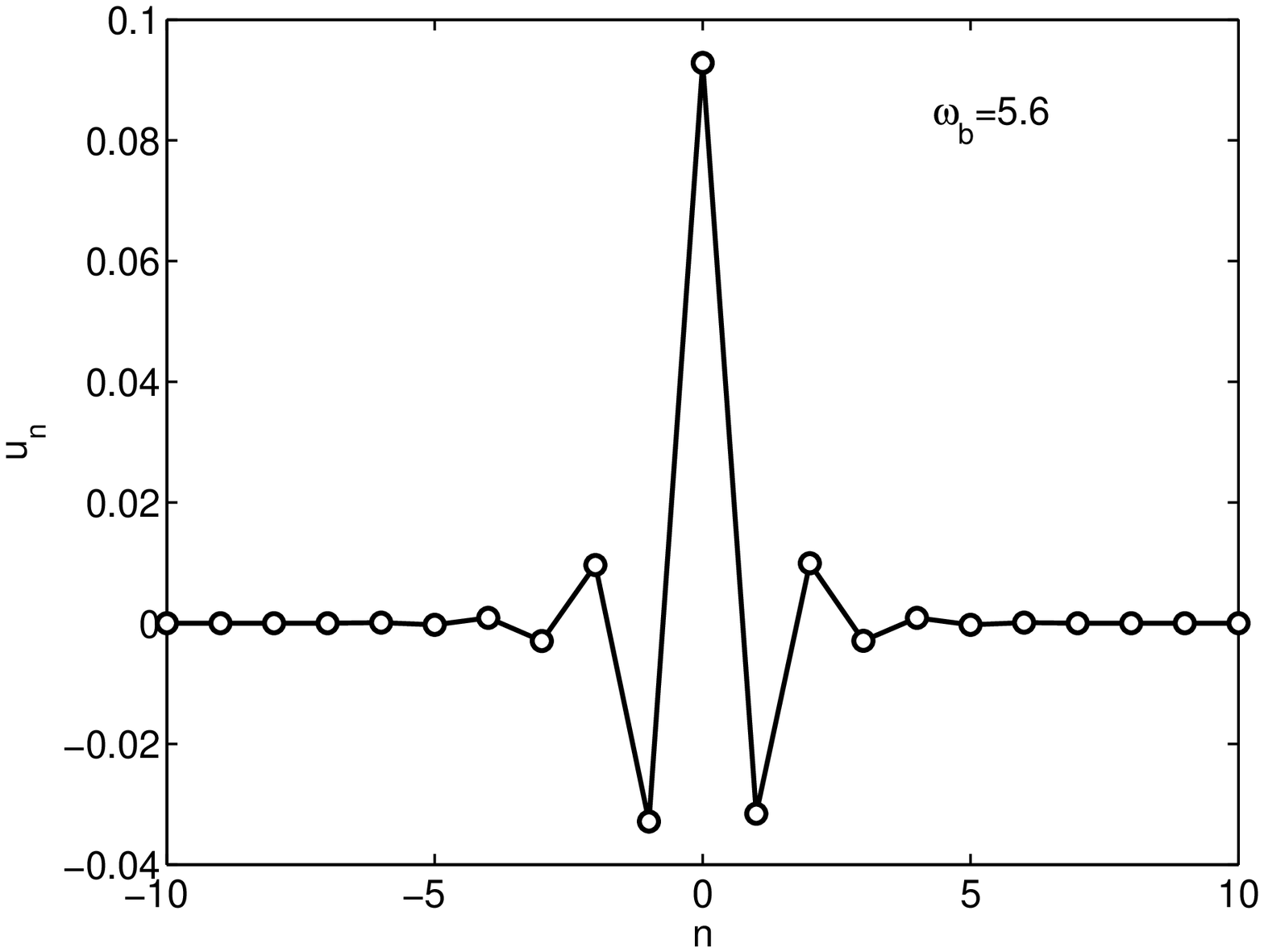}\\
\includegraphics[width=0.8\columnwidth]{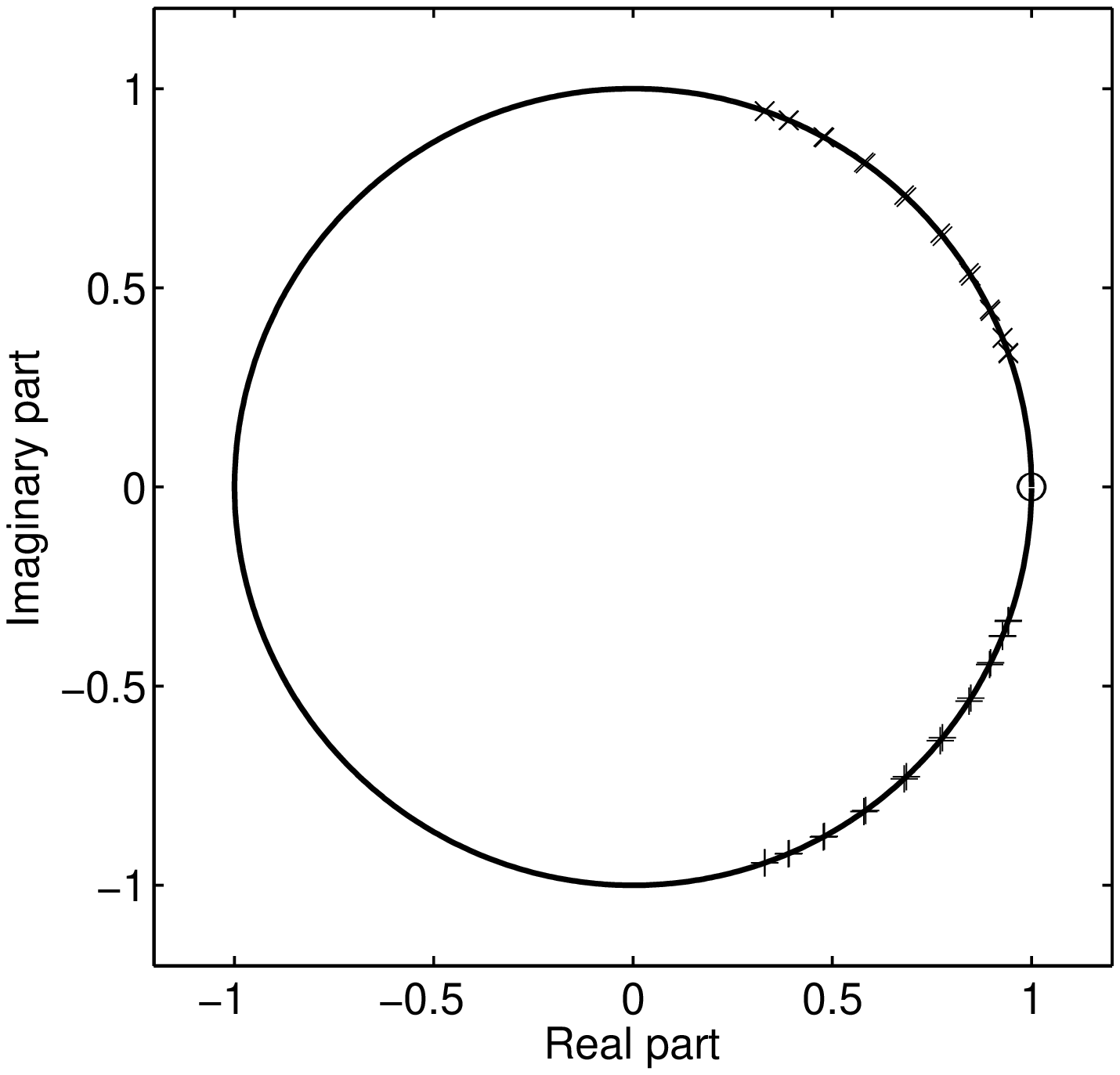}
\end{center}
\caption{{\bf Top}: Profile of the site-centered breather with lowest frequency $\omega_b=5.6$, slightly above the top of the phonon band.  {\bf Bottom}: Eigenvalues of the Floquet matrix for the same breather, showing its high degree of linear stability. The breathers becomes more localized as the frequency increases and the Floquet eigenvalues move away from unity. The asymmetric bond centered mode is however always unstable and exists until the frequency is close to twice the top of the phonon band.
}
\label{fig_breather}
\end{figure}

%juan: jul1: reasoning about the lack of moving vacancies with three references added.
The properties of the repulsive interaction potential which corresponds to our physical model imply that it is very steep when the ions approach but there is no restoring force when the bonds are stretched. For kinks with $Q=1$, i.e., moving vacancies or rarefaction solitary waves, to exist in a given system, it is necessary a steep potential corresponding to the stretching of the bonds~\cite{forbes2012,kosevich2017}, except in systems with some special dispersion relations~\cite{herbold2013}. This implies that moving vacancies are not possible in our system, which has been confirmed by extensive numerical search for kinks with topological charge $Q=1$ with negative results.
%YURIY: look for a reference?

\section{Breathers}
\label{sec:breathers}
In this section, we describe first stationary breathers, their stabilities and energies and, second, the small amplitude moving breathers that we have found in the system.
\subsection{Stationary breathers}

We construct stationary breathers
starting from the anticontinuous limit\,\cite{mackayaubry94,marinaubry96}, that is, initially setting $c_s$=0, so as to prevent the coupling between particles, and obtaining the solutions
for the isolated oscillator:
\begin{equation}
\ddot{u}_n=-V'(u_n),\,
\label{eq:ddotun1}
\end{equation}
where $n$, can be any of the oscillators, but we will use $n=0$. This study is also interesting to better understand the properties of the substrate potential and
 the system as a whole. The potential profile and the frequency of a single oscillator as a function of the amplitude is represented in Fig.~\ref{fig:wA_single}.

%%%%%%%%%%%%%%%%%%%%%%%%%%%%%%figure12
\begin{figure}[htb]
\begin{center}
\includegraphics[width=0.9\columnwidth]{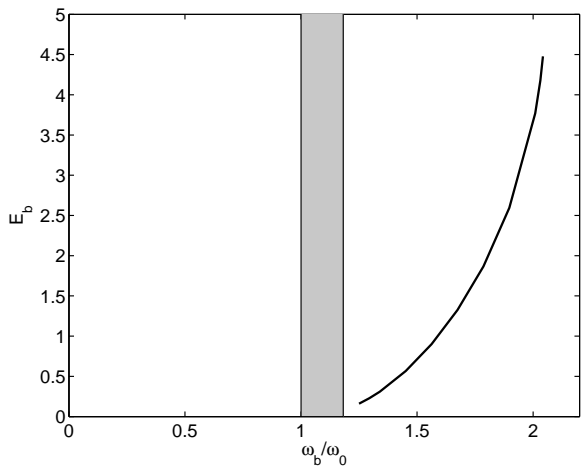}%0.6\textwidth]
\end{center}
\caption{Energy of the site-centered breathers in scaled units ($u_E$=2.77\,eV) with respect to their frequency normalized to the isolated oscillator frequency $\omega_0$. The shaded area marks the phonon frequencies. The energy range from 0.16 (0.44\,eV) to 4.47 (12.4\,eV). Bond-centered breathers can have more energy but they are unstable.}
\label{fig_breatherenergy}
\end{figure}

%%%%%%%%%%%%%%%%%%%%%%%%%figure1
\begin{figure}[b]
\begin{center}
\includegraphics[width=0.9\columnwidth]{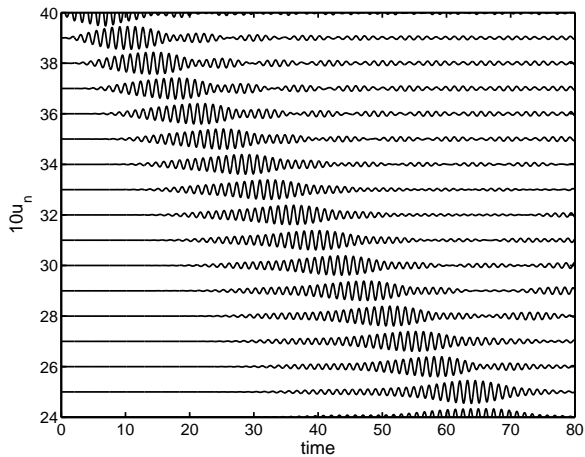}%0.6\textwidth
\caption{Moving breather produced by an asymmetric compression of two particles with $u_n(0)=-0.15$ and $u_{n-1}(0)=-0.1$ , which produce a large stationary breather and a small amplitude moving breather. It propagates smoothly rebounding in the initial perturbation several thousand units of time surviving to the phonons produced in the system by the perturbation.\label{fig:movingbreather}
}
\end{center}
\end{figure}
It can be seen that differently from a sinusoidal potential, the substrate potential is initially hard, because the frequency of the nonlinear oscillator increases with the amplitude.

By a continuation method, increasing the coupling at small intervals, it is possible to obtain breathers from a frequency close to the top of the phonon band to the resonance frequency $2\omega_0$ with the second harmonic of the lowest frequency of the phonon band.  Breathers are typical mode-$\pi$ with a staggered profile, both site-centered (Sievers Takeno or ST)\cite{sievers-takeno1988} or bond-centered (Page mode or P)\cite{page1990}. The ST-breather  can be seen in Fig.~\ref{fig_breather} for a very low frequency and it is stable. It exists until its frequency is about twice the lowest phonon frequency $\omega_0$. The P breather is always unstable and it exists until its frequency is about twice the top of the phonon band.
They become more and more localized as the frequency increases. As can be seen in Fig.~\ref{fig_breatherenergy}, the energy of the ST-breather stretches from 0.16 (0.44\,eV) to 4.47 (12.4\,eV) or about half the energy of the sliding kinks, the P-breather having about the double of energy although they coincide close to the phonon band. Breathers can be produced by $^{40}$K recoil as $\beta^-$ decay has a continuous range of energies from 0 to 42\,eV.

\subsection{Moving breathers}
\label{ssec:mv_breathers}
Breathers can be moved with several methods, for that it is necessary that the breather is not extremely stable and a pair of Floquet eigenvalues are close to the unity, meaning that other solution is close. Then, a perturbation with the momenta of the eigenvector corresponding to the instability eigenvalue of a localized mode can bring about breather movability. This is not the case, as the eigenvalues can clearly be identified with the phonon band $\exp(\text{i}\omega_\text{phon}T_b)=\exp(\text{i}2\pi\omega_\text{phon}/\omega_b)$  without any pinning mode. Numerical attempts to
perturb the breathers with many variants used usually in the literature as the discrete gradient and others have failed. Also the introduction of the expansion of the interaction to more neighbours which has been shown to enhance mobility has not been successful.

However, we have observed the smooth movement of small amplitude and energy breathers produced by some simple but meaningful initial conditions as compressing two particles with some asymmetry. A large stationary breather is produced but also a small moving breather that survives more than one thousand periods even rebounding into the large stationary breather. The velocities and energies of the moving breathers are small, a typical example shown in Fig.~\ref{fig:movingbreather}
has $s\simeq 0.3, E\simeq 0.05\simeq 0.15$\,eV. Its velocity is 10 times smaller and its energy 150  times smaller that the single crowdion energy. This is an interesting result as there are many tracks of different types scattering from the primary tracks in muscovite mica. The perturbation that produce them needs to have a much smaller energy.  A more detailed study is underway and it will be published in due time. Note that small-amplitude moving envelope solitons and their merging have been described on the FPU chain\,\cite{kosevich-lepri2000}.

\section{Conclusions}
\label{sec:conclusions}
In this article we have investigated the existence of localized nonlinear waves in a model for the potassium ions in muscovite mica with potentials obtained from physical properties and empirical potentials fitted for other properties of the material or for short range collisions. The motivation is the fossil and experimental evidence that nonlinear excitations can travel long distances along the closed packed lines of the potassium layer.
We have performed a systematic study of the travelling waves in this system and we have found a wide range of nanopterons and non-radiating kinks. The only stable ones being a single crowdion and two double crowdions with velocities about five times the sound velocity.  Nanopterons of small tail amplitude can be even more stable than the sliding kinks. A necessary condition for kinks and nanopterons stability is that their velocity corresponds to a phonon wavenumber in the first Brillouin zone.  We have also constructed stationary breathers that cannot be moved, but that can be of interest for other properties of the crystal. There are, also, small amplitude breathers that can be produced by simple physical conditions, with much smaller energies than  the crowdions and bi-crowdions and with a long survival time.   Among the nonlinear waves that we have found, only the single crowdion had been described  previously with a less rigorous method. Crowdions and bi-crowdions are also natural charge carriers in an ionic crystal. Nanopterons with small amplitude tails and high stability may play an important role as long-lived transient states. An interesting subject for future work is the study of charge transport along the lattice close-packed lines as electron or hole hopping coupled with some lattice excitations. We are working in a tight-binding model with these characteristics which will be published in due time. We adventure the hypothesis that the high energy single and double crowdions could be the primary charge carriers along lattice lines, but not of the secondary charge carriers that are scattered by it, which could be breathers coupled to a charge.

\section*{Acknowledgments}
JFRA wishes to thank project FIS2015-65998-C2-2-P from MINECO and grant 2017/FQM-280  from Junta de Andalucia, Spain. He also acknowledges Prof. Yusuke Doi and the University of Osaka for hospitality. JFRA and YuAK acknowledge grants from VI PPIT-US of the Universidad de Sevilla, Spain, which funded research stays in Osaka and Sevilla, respectively.
 YZ acknowledges the support of the National Academy of Sciences of
Ukraine through program No. 0117U000236. YuAK acknowledges funding from the Federal Agency of Scientific Organizations of Russia (Research Topic 0082-2014-0013, No. AAAA-A17-117042510268-5). YD  acknowledges the supports of JSPS KAKENHI Grant Number 16K05041.
%\bibliography{bound}
%\end{document}
%merlin.mbs aipnum4-1.bst 2010-07-25 4.21a (PWD, AO, DPC) hacked
%Control: key (0)
%Control: author (8) initials jnrlst
%Control: editor formatted (1) identically to author
%Control: production of article title (0) allowed
%Control: page (1) range
%Control: year (1) truncated
%Control: production of eprint (0) enabled
\newcommand{\noopsort}[1]{} \newcommand{\printfirst}[2]{#1}
  \newcommand{\singleletter}[1]{#1} \newcommand{\switchargs}[2]{#2#1}

\end{document}